\documentclass{article}

\PassOptionsToPackage{numbers, compress}{natbib}
\usepackage[preprint]{neurips_2026}

\usepackage[utf8]{inputenc} 
\usepackage[T1]{fontenc}    
\usepackage{hyperref}       
\usepackage{url}            
\usepackage{booktabs}       
\usepackage{amsfonts}       
\usepackage{nicefrac}       
\usepackage{microtype}      
\usepackage{xcolor}         

\usepackage{amsmath}
\usepackage{graphicx}
\usepackage{booktabs}
\usepackage{multirow}
\usepackage[table]{xcolor}
\usepackage{xspace}
\usepackage{subcaption}
\usepackage{adjustbox}

\definecolor{dustyblue}{RGB}{220, 235, 245}
\definecolor{sagegreen}{RGB}{225, 235, 225}
\definecolor{mistyorange}{RGB}{245, 230, 220}
\definecolor{softpurple}{RGB}{235, 230, 245}

\newcommand{\headerblue}{\rowcolor{dustyblue}}
\newcommand{\headergreen}{\rowcolor{sagegreen}}
\newcommand{\headercolorlong}{\rowcolor{gray!17}}

\newcommand{\promptheader}[1]{%
    \colorbox{gray!17}{\parbox{\dimexpr\linewidth-2\fboxsep\relax}{\textbf{#1}}}%
}

\newcommand{\tool}{\textsc{Tracer}\xspace}

\title{\tool: A Semantic-Aware Framework for Fine-Grained Contamination Detection in Code LLMs}

\author{%
  Yifeng Di \\
  Purdue University\\
  West Lafayette, IN \\
  \texttt{di5@purdue.edu} \\
  \And
  Xuliang Huang \\
  University of Illinois Urbana-Champaign\\
  Champaign, IL \\
  \texttt{xh54@illinois.edu} \\
  \And
  Tianyi Zhang \\
  Purdue University\\
  West Lafayette, IN \\
  \texttt{tianyi@purdue.edu} \\
}

\begin{document}

\maketitle

\begin{abstract}

Data contamination is a known threat to the reliability of model evaluation. However, it remains underexplored in code large language models (LLMs), where contamination often goes beyond exact duplication. 
We present \tool, a semantic-aware framework for fine-grained code contamination detection. \tool models contamination using three levels of semantic overlap---\emph{Functionally Identical}, \emph{Nearly Identical}, and \emph{Shared Logic}---and detects them through a coarse-to-fine pipeline. 
We also introduce the first benchmark for fine-grained code contamination detection, spanning three widely used benchmarks and three representative post-training datasets. 
\tool achieves strong and consistent performance across multiple LLM backbones, with \textsc{GPT-5} reaching an F1 score of 0.91 in fine-grained detection. In the binary setting, \tool attains an F1 of 0.92, outperforming existing methods by 42\%–217\%. We further conduct ablation studies and error analysis to assess the contributions of individual components in \tool.

\end{abstract}

\section{Introduction}

Data contamination poses a significant threat to the reliability of model evaluation~\citep{xu2024benchmark, dong2024generalization, balloccu2024leak, deng2024investigating}, yet it remains underexplored in Code Large Language Models (Code LLMs). Code LLMs are routinely evaluated on public benchmarks, such as HumanEval~\citep{chen2021evaluating}, MBPP~\citep{austin2021program}, and LiveCodeBench~\citep{jain2025livecodebench}. 
However, such evaluation assumes that benchmark tasks do not already appear, in exact or similar form, in the data used for training or fine-tuning. When this assumption is violated, the evaluation scores may reflect memorization rather than generalization, leading to overly optimistic conclusions about model capability. This risk is especially salient in code, where benchmarks and post-training corpora often draw or adapt from overlapping public sources such as tutorials, programming competitions, and GitHub repositories.

In practice, a benchmark coding task may reappear in post-training data with paraphrased instructions, slightly modified constraints, or different input-output formats while preserving the same underlying computational objective. However, existing contamination detection methods are typically built around a binary notion of data overlap and commonly rely on string matching~\citep{chowdhery2023palm, achiam2023gpt, touvron2023llama}, heuristic similarity thresholds~\citep{lee2023platypus, piktus2023roots, riddell2024quantifying}, or LLM judgments~\citep{witteveen2019paraphrasing, abaskohi2023lm, yang2023rethinking}. While effective for identifying near duplicates, these methods provide limited support for distinguishing the different levels of semantic overlap in coding benchmarks.

We therefore formulate code contamination detection as a \emph{fine-grained semantic categorization} problem rather than a binary matching problem. Inspired by code clone detection~\citep{roy2007survey, svajlenko2015evaluating, sajnani2016sourcerercc, ain2019systematic}, which moves beyond exact duplication to distinguish different levels of syntactic and semantic similarity, we introduce three types of code contamination: \emph{Functionally Identical}, \emph{Nearly Identical}, and \emph{Shared Logic}. This formulation distinguishes severe leakage from weaker semantic transfer, providing a more informative basis for analyzing how contamination may inflate benchmark performance.

Based on this formulation, we propose \tool (\textbf{T}ask-level \textbf{R}etrieval-\textbf{A}ugmented 
\textbf{C}ontamination \textbf{E}valuation with \textbf{R}easoning), a semantic-aware framework for fine-grained contamination detection. Given a post-training dataset and a coding evaluation benchmark, \tool first normalizes task descriptions to reduce superficial formatting and phrasing variation between these two datasets. Second, it applies embedding-based triage to all task pairs between the two datasets to narrow the candidate space. For each candidate pair, an LLM backbone assigns a fine-grained contamination label based on task intent, constraints, and underlying algorithmic logic. Finally, it filters out trivial programming tasks (e.g., basic arithmetic, simple list operations, or standard I/O formatting) to avoid overstating contamination. This design focuses analysis on semantically meaningful overlap that is more likely to threaten evaluation validity.

We also curate the first dataset for fine-grained code contamination detection, which includes 540 manually annotated benchmark--training task pairs spanning three widely used code generation benchmarks and three representative post-training corpora. 
On fine-grained contamination detection, \tool achieves strong and stable performance across four LLM backbones, with \textsc{GPT-5} reaching the best overall F1 of 0.91 while weaker models remain competitive. Since existing contamination detection methods only support binary detection, we further compare \tool with existing methods in the binary setting. \tool achieves 0.92 F1, substantially outperforming three baselines by 42\%-217\%. Finally, we perform ablation studies and error analysis to examine the contribution of individual stages and the remaining challenges in contamination detection in Code LLMs.

\section{Related Work}

\subsection{Data Contamination Detection}
\label{sec:contam}

Prior work on data contamination detection can be categorized into two settings: (1) methods that directly compare evaluation instances against accessible training corpora, and (2) methods that infer contamination from model behavior when the training data are unavailable~\citep{ravaut2024comprehensive, cheng2025survey, xu2024benchmark}.

When the training data are available, a commonly used method is lexical matching, which is highly scalable to large text corpora~\citep {brown2020language, chowdhery2023palm, achiam2023gpt, touvron2023llama, dekoninck2024evading}. Another line of work uses semantic similarity, where embedding-based or hybrid similarity improves recall beyond lexical matching~\citep{lee2023platypus, gunasekar2023textbooks, riddell2024quantifying}. More recently, some work uses LLM-based semantic verification, which is more robust to paraphrasing~\citep{witteveen2019paraphrasing, abaskohi2023lm, yang2023rethinking}. All existing methods treat contamination detection as a binary classification problem. By contrast, in this work, we introduce a fine-grained categorization of code contamination that captures the degrees of semantic information leakage from an evaluation benchmark to training data. Nevertheless, we demonstrate that {\tool} still achieves significantly better performance than these methods in the binary setting (Section~\ref{sec:categorization_accuracy}).

When the training data are not available, prior work infers contamination based on performance anomalies~\citep{zhao2024cap}, membership inference~\citep{duan2024membership}, and likelihood-based confidence analysis~\citep{shi2023detecting, zhang2024min, dong2024generalization}. These methods typically assume that contaminated data induce detectable memorization signals in model behavior. However, they often struggle to separate contamination from generalization~\citep{fu2025does}. In this work, we focus on the setting where training data are available, which enables direct comparison between benchmark and training instances.

\subsection{Contamination-Free Benchmark}

A parallel line of work addresses data contamination by constructing contamination-free benchmarks~\citep{cheng2025survey, xu2024benchmark}. 
One common strategy is to limit the exposure of evaluation data to public training corpora~\citep{rajore2024truce, jacovi2023stop}. 
Another direction uses dynamic benchmarks, which continuously update or newly collect evaluation instances to stay ahead of training cutoffs 
\cite{li2024latesteval, white2024livebench, jain2025livecodebench}. 
Furthermore, some work cleans existing benchmarks through rewriting, regeneration, or similarity-based filtering to remove potentially contaminated instances~\citep{zhu2024clean, deng2024investigating, li2023cleva, yang2023rethinking}. This line of work focuses on benchmark curation and cleaning, rather than contamination detection itself. 

\begin{table*}[t]
\centering
\small
\caption{Illustrative examples of the fine-grained code contamination categories defined in Section~\ref{sec:problem_formulation}.}
\label{tab:contamination_examples}
\resizebox{\textwidth}{!}{
\begin{tabular}{l c p{5.7cm} p{5.7cm}}
\toprule
\textbf{Contamination Category} & \textbf{Abbr.} & \textbf{Benchmark Task Example} & \textbf{Training Task Example} \\
\midrule
Functionally Identical & \textbf{FI}
& Given a list of integers, return the maximum element.
& Write a function that finds the largest number in an array. \\
\midrule
Nearly Identical & \textbf{NI}
& Given a list of integers, return the maximum element.
& Given a non-empty list of integers, return the maximum value and its index. \\
\midrule
Shared Logic & \textbf{SL}
& Implement a function that takes a list of strings and returns the string with the maximum number of characters. If there are multiple such strings, break ties by lexicographical order.
& Given a list of strings, write a function to print the longest string. \\
\midrule
Unrelated & \textbf{U}
& Reverse a string.
& Compute the shortest path in a weighted graph using Dijkstra's algorithm. \\
\bottomrule
\end{tabular}
}
\end{table*}

\subsection{Code Clone Detection}

Code clone detection is closely related to our problem because data contamination in code benchmarks often involve highly similar programming content. The code clone literature  distinguishes four types of code clones, from exact copies to semantic clones, which inspires our taxonomy of code contamination~\citep{kamiya2002ccfinder, roy2007survey, zakeri2023systematic, li2017cclearner}. Existing methods detect clones using text similarity~\citep{johnson1994substring}, token similarity~\citep{kamiya2002ccfinder, roy2008nicad, sajnani2016sourcerercc}, tree similarity~\citep{jiang2007deckard}, or more expensive semantic techniques based on program dependence~\citep{krinke2001identifying, gabel2008scalable} or learned representations~\citep{li2017cclearner, white2016deep}.

However, clone detection does not directly solve our task. Our goal is to compare benchmark and training \emph{tasks}, which are primarily specified in natural language descriptions. Moreover, the same programming task may involve multiple correct solutions with little syntactic similarity, making code-level clone signals weak or absent even when task-level overlap is strong. 
This is confirmed in our comparison against a recent program-similarity-based baseline (Section~\ref{sec:categorization_accuracy}).

\section{Problem Formulation}
\label{sec:problem_formulation}

Unlike existing work, which treats data contamination detection as a binary classification problem, we introduce a fine-grained categorization of code contamination that captures different degrees of semantic information leakage from an evaluation benchmark to training data.

Let $D_{\text{train}}$ denote a post-training dataset for fine-tuning a pre-trained LLM for coding tasks, and let $D_{\text{test}}$ denote an evaluation benchmark. We define a label set $\mathcal{Y}$ that captures different degrees of semantic relatedness between a task in $D_{\text{train}}$ and a task in $D_{\text{test}}$:
\begin{equation}
    \mathcal{Y} = \{ \text{FI}, \text{NI}, \text{SL}, \text{U} \}.
\end{equation}

We define each category below: 
\begin{itemize}
    \item \textbf{Functionally Identical (FI):} The two tasks specify the same computational objective with equivalent inputs and outputs, such that their solutions are interchangeable.
    \item \textbf{Nearly Identical (NI):} One task is a minor variant of the other, differing only in constraints, formatting, or auxiliary requirements while solving the same underlying problem.
    \item \textbf{Shared Logic (SL):} The tasks address different objectives but rely on the same core algorithmic or reasoning strategy, enabling transfer of problem-solving logic.
    \item \textbf{Unrelated (U):} The tasks share no meaningful semantic, algorithmic, or functional overlap.
\end{itemize}

To further clarify the distinctions between these categories, Table~\ref{tab:contamination_examples} provides illustrative examples of task pairs corresponding to each category. 

Under this formulation, code contamination detection is defined as the problem of assigning each candidate task pair to one category in $\mathcal{Y}$. Formally, we seek to learn or approximate a mapping function
\begin{equation}
    f: D_{\text{train}} \times D_{\text{test}} \rightarrow \mathcal{Y}.
\end{equation}

Because the Cartesian product $|D_{\text{train}}| \times |D_{\text{test}}|$ is prohibitively large in practice, any instantiation of $f$ must incorporate an efficient mechanism to filter high-probability candidate pairs prior to fine-grained semantic classification.

\section{Approach}
\label{sec:approach}

\begin{figure*}[t]
\centering
\includegraphics[width=\linewidth]{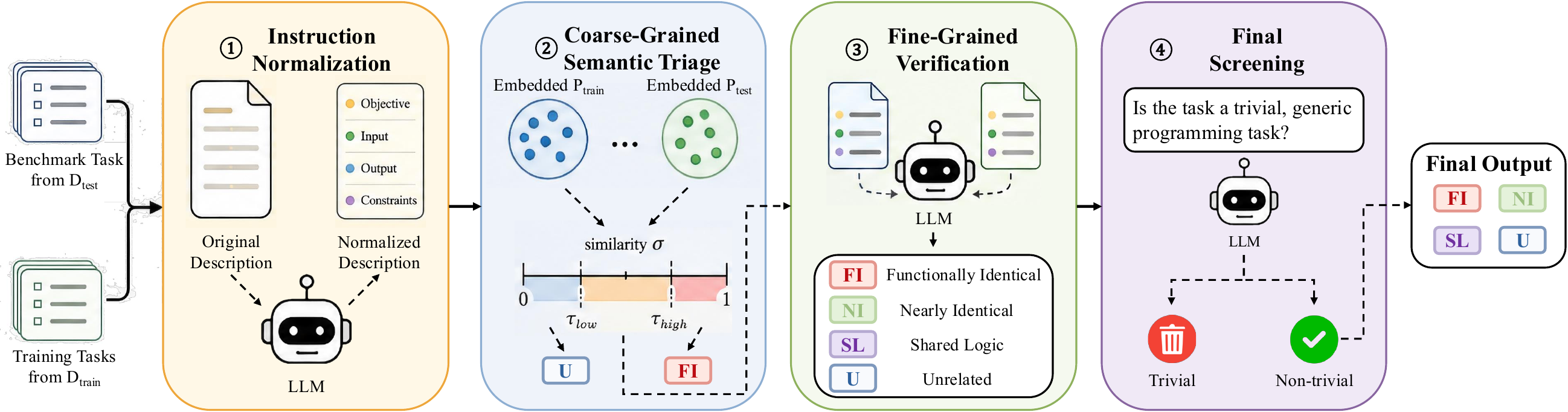}
\caption{An Overview of \tool.}
\label{fig:overview}
\end{figure*}

We propose \tool, a semantic-aware framework for instantiating the contamination categorization function 
\( f: D_{\text{train}} \times D_{\text{test}} \rightarrow \mathcal{Y} \) in practice. 
The central challenge is balancing semantic precision with computational efficiency over a combinatorial task-pair space.

To address this challenge, we adopt a coarse-to-fine design that progressively refines semantic certainty while reducing the number of task pairs that require expensive semantic verification. Figure~\ref{fig:overview} illustrates the overall pipeline. For each task from $D_{\text{test}}$, \tool compares it against every single task in $D_{\text{train}}$  through four stages. It first normalizes task descriptions to reduce superficial variation in formatting and phrasing. It then applies embedding-based triage to discard clearly unrelated pairs and retain only candidate pairs with potential semantic overlap. Next, it uses LLM-based verification to assign a fine-grained label from $\mathcal{Y}=\{\text{FI}, \text{NI}, \text{SL}, \text{U}\}$. Finally, it applies trivial-task filtering to exclude generic utility-style tasks that may otherwise overstate contamination.

\subsection{Instruction Normalization}
\label{sec:normalization}

Before comparing tasks for potential overlap, we first address superficial variation in natural language descriptions. Task descriptions from different datasets are often written in different formats, conventions, or levels of verbosity, even when they describe closely related programming objectives. Such inconsistencies can obscure semantic overlap and introduce noise into similarity measurements.

To reduce this variability, we apply an instruction normalization step that rewrites each task description into a standardized format that preserves the task semantics while making its key components explicit. Given an original description \( p \), \tool prompts an LLM to produce a reformatted description \( \tilde{p} \) that follows a consistent structural template, including the problem objective, input specification, output requirements, and relevant constraints. Table~\ref{tab:normalization-prompt} from Appendix~\ref{sec:prompts} describes the prompt used for this normalization step. To assess whether normalization changes task semantics, we further conduct a manual analysis on task descriptions sampled from all six datasets used in our evaluation. Specifically, we randomly sample 20 task descriptions from each of the three evaluation benchmarks and three post-training corpora, resulting in 120 sampled task descriptions in total. We normalize each sampled task using the four LLM backbones, producing 480 original-normalized task pairs for manual inspection. The results, reported in Appendix~\ref{app:normalization-analysis}, show that normalization preserves the original task semantics for 95\%--100\% of the pairs across the four LLM backbones and that the manual labels exhibit high inter-annotator agreement ($\kappa = 0.92$).

Appendix~\ref{sec:normalization-example} provides an example in which two differently structured descriptions of the same programming task, drawn from an evaluation benchmark and a post-training corpus and later categorized as Functionally Identical, are rewritten into a uniform format. As illustrated, the normalization step primarily removes extraneous stylistic variation and reduces formatting inconsistencies across post-training corpora, which improves the comparability of tasks from different datasets for the following stages.

\subsection{Coarse-Grained Semantic Triage}
\label{sec:triage}

Exhaustively applying fine-grained LLM verification to every benchmark--training task pair is computationally prohibitive for large post-training corpora. For example, the Cartesian product $|D_{\text{train}}| \times |D_{\text{test}}|$ easily reaches more than two million in our evaluation settings.

\tool uses embedding-based triage as a screening stage to reduce the number of pairs that require downstream LLM-based verification. It performs a high-throughput, cost-aware triage based on embedding similarity, with the goal of rapidly filtering out obviously unrelated pairs while retaining high recall for potentially contaminated cases. 

After task normalization, each task description \( \tilde{p} \) is embedded into a fixed-dimensional vector space using a text embedding model. For each benchmark task description from $D_{\text{test}}$, we compute similarity scores $\sigma \in [0,1]$ against each post-training task description from $D_{\text{train}}$. Based on the similarity score, we then partition pairs into three regions using a lower screening threshold $\tau_{\mathrm{low}}$ and an upper screening threshold $\tau_{\mathrm{high}}$. Pairs with $\sigma \geq \tau_{\text{high}}$ are considered exact duplicates and are directly labeled as \text{FI}. Pairs with $\sigma \leq \tau_{\text{low}}$ are unlikely to reflect meaningful semantic overlap and are therefore labeled as \text{U}. Pairs with similarity scores between these thresholds fall into an \emph{ambiguity zone}, where embedding similarity alone is insufficient to determine their semantic relationship. These pairs are subsequently passed to the next stage for fine-grained verification.

\subsection{Fine-Grained Verification}
\label{sec:fine-grained-verification}

For pairs with similarity scores between $\tau_{\mathrm{low}}$ and $\tau_{\mathrm{high}}$ from the previous stage, \tool performs fine-grained verification using an LLM backbone. Given the normalized task descriptions, the LLM backbone is prompted to assess their semantic similarity and categorize it based on the definitions of the four categories in \( \mathcal{Y} \). 
Table~\ref{tab:verification-prompt} from Appendix~\ref{sec:prompts} presents the full prompt template used in our experiments. In particular, it instructs the model to specify the relationship between the two tasks by comparing their objectives, constraints, input--output behavior, and underlying algorithmic logic, producing a forced-choice decision consistent with the formulation in Section~\ref{sec:problem_formulation}.

\subsection{Final Screening}
\label{sec:final-screening}

Even when two tasks share similar semantics, they may correspond to trivial programming tasks that only involve basic or atomic operations, e.g., computing sums or averages, simple list manipulations, or standard IO operations. These basic tasks frequently appear across tutorials, textbooks, and beginner programming exercises. Because these tasks are widely reused throughout various programming tasks, their presence in both benchmark and post-training corpora does not necessarily indicate meaningful information leakage. Counting such cases as contamination can therefore lead to an overestimation of contamination rates.

To address this issue, \tool performs a final screening step to identify and exclude trivial tasks whose computational objectives are generic and broadly shared across programming materials. This design is motivated by prior work~\citep{roy2007survey, zhang2017automated, li2017cclearner} in code clone detection that filters out uninteresting or trivial clones to focus analysis on more meaningful forms of code similarity. Similarly, our goal is to focus contamination detection on tasks whose computational objectives are specific and non-trivial, rather than generic programming constructs that appear pervasively across datasets. Specifically, \tool prompts an LLM to determine whether a task only involves basic or atomic operations. Table~\ref{tab:final-screening-prompt} in Appendix~\ref{sec:prompts} shows the prompt used in this stage. If one task from a task pair is classified as trivial, the task pair is excluded from contamination reporting.

\section{Experiments}
\label{sec:Evaluation}

\subsection{Benchmark Construction}
\label{sec:golden_dataset}

To evaluate fine-grained contamination detection, we construct a manually annotated benchmark of coding task pairs drawn from three evaluation benchmarks (\textsc{HumanEval}~\citep{chen2021evaluating}, \textsc{MBPP}~\citep{austin2021program}, and \textsc{LiveCodeBench-v6}~\citep{jain2025livecodebench}) and three code post-training corpora (\textsc{CodeAlpaca-20k}~\citep{codealpaca}, \textsc{Evol-CodeAlpaca-V1}~\citep{luo2023wizardcoder}, and \textsc{Magicoder-OSS-Instruct-75K}~\citep{wei2023magicoder}), yielding nine benchmark-training combinations. Because the full Cartesian product is extremely large and contamination cases are rare, we first use embedding-based retrieval to identify candidate pairs and then manually annotate a sampled subset.

Specifically, we compute similarity scores between benchmark tasks and post-training tasks using \texttt{jina-embeddings-v3}. We retain a task pair as a candidate if its similarity score exceeds 0.5, yielding a candidate pool of 171,654 task pairs. From this pool, we randomly sample 85 pairs from each of the nine benchmark-training combinations, resulting in 765 sampled pairs in total.

We then split the sampled pairs into a development set of 225 pairs and a held-out test set of 540 pairs. For each benchmark-training combination, this corresponds to 25 development pairs and 60 test pairs. The development split is used for hyperparameter tuning for all evaluated methods, including the similarity thresholds used by \tool and the thresholds used in  baselines. All final evaluation results reported are computed on the held-out test split. Each pair is independently annotated by two authors, achieving strong inter-annotator agreement (Cohen’s $\kappa = 0.81$). All disagreements were subsequently resolved through discussion. Table~\ref{tab:golden_distribution} reports the full label distribution. Additional details on annotation are provided in Appendix~\ref{app:benchmark_construction}.

\begin{table}[t]
\centering
\small
\caption{Category distribution in the curated benchmark.}
\label{tab:golden_distribution}
\begin{tabular}{lccc}
\toprule
\textbf{Category} & \textbf{Abbr.} & \textbf{Count} & \textbf{Percentage} \\
\midrule
Functionally Identical & FI & 29  & 5.4\% \\
Nearly Identical       & NI & 108 & 20.0\% \\
Shared Logic           & SL & 128 & 23.7\% \\
Unrelated              & U  & 275 & 50.9\% \\
\bottomrule
\end{tabular}
\end{table}

\subsection{Experimental Design}
\label{sec:experimental-design}

We organize the evaluation around two settings. The primary setting is \textit{fine-grained contamination detection}, which evaluates whether a method can distinguish among the semantic contamination categories defined in Section~\ref{sec:problem_formulation}. The secondary setting is \textit{binary contamination detection}, introduced only to enable comparison with existing methods since they do not support fine-grained categorization.

\paragraph{Evaluation on Fine-grained Contamination Detection.}
We first evaluate \tool on the held-out set of the benchmark introduced in Section~\ref{sec:golden_dataset}. Since there is no prior benchmark or directly comparable method for fine-grained code contamination detection, this setting focuses on comparing different LLM backbones within our framework. We use \textsc{jina-embeddings-v3}~\citep{sturua2024jina} as the embedding model for semantic triage and evaluate multiple LLM backbones, including \textsc{GPT-5}, \textsc{Gemini-2.5-Pro}, \textsc{gpt-oss-120b}, and \textsc{Qwen3-14B}. Complete model details are provided in Appendix~\ref{app:models}.

In this setting, we report category-level precision, recall, and F1 for the four classification labels, together with \emph{macro}-averaged precision, recall, and F1 across all labeled pairs. 

\paragraph{Evaluation on Binary Contamination Detection.}
We also evaluate a binary reduction of the task in order to compare \tool with existing contamination detection methods. In this setting, \emph{Functionally Identical}, \emph{Nearly Identical}, and \emph{Shared Logic} are grouped as \emph{contaminated}, while \emph{Unrelated} is treated as \emph{non-contaminated}. 
We compare \tool against three baselines that cover the three representative types of research in data contamination detection when training data is available, as discussed in Section~\ref{sec:contam}. First, \textsc{ROOTS}~\citep{piktus2023roots} serves as a lexical matching baseline that captures near-exact or surface-form overlap. Following prior work, we use a sparse BM25 index implemented with Pyserini~\citep{lin2021pyserini} and native Lucene analyzers\footnote{\label{note1}\href{https://lucene.apache.org/}{https://lucene.apache.org/}}to score lexical relevance between paired tasks, and classify a task pair as contaminated when the BM25 score exceeds a decision threshold. 
Second, Riddell et al.~\citep{riddell2024quantifying} propose a contamination quantification pipeline that integrates surface-level textual similarity with semantic program similarity. As this pipeline produces a similarity score rather than a binary classification, we adapt it by introducing a threshold to obtain binary decisions. We refer to this baseline as \textsc{Program Similarity Matching} (\textsc{PSM}).  Third, \textsc{LLM Decontaminator}~\citep{yang2023rethinking} adopts a question-answering formulation, using an LLM backbone to determine whether a benchmark task should be considered contaminated. For a fair comparison, we use GPT-5 as the LLM backbone for both \tool and \textsc{LLM Decontaminator} in this evaluation setting. 
We report binary precision, recall, and F1 in this setting.

\paragraph{Threshold Selection.} \tool, \textsc{ROOTS}, and \textsc{PSM} all rely on thresholds to classify task pairs. To be consistent, we tune all of them on the same development split introduced in Section~\ref{sec:golden_dataset}. For \tool, embedding-based triage uses two screening thresholds: a lower threshold $\tau_{\mathrm{low}}$ and an upper threshold $\tau_{\mathrm{high}}$. We tune both on the development split via grid search with step size 0.05. The lower threshold is chosen to preserve high recall over contaminated pairs, reducing the chance that true contamination cases are pruned before LLM-based verification, while the upper threshold is chosen to achieve high precision among strong semantic candidates so that pairs above it correspond to Functionally Identical matches. This yields $\tau_{\mathrm{low}} = 0.6$ and $\tau_{\mathrm{high}} = 0.9$, with the lower threshold preserving recall above 0.9 and the upper threshold achieving precision above 0.9 on the development split. For \textsc{ROOTS} and \textsc{PSM}, we tune the BM25 and program-similarity thresholds on the same development split by maximizing binary F1, yielding thresholds of 0.6 and 0.65, respectively. Appendix~\ref{sec:threshold} reports all the precision--recall curves of threshold tuning.

\subsection{Results and Analysis}
\label{sec:categorization_accuracy}

\paragraph{Fine-grained Contamination Detection.}

\begin{table*}[t]
\centering
\small
\caption{Fine-grained contamination detection results across benchmarks and post-training datasets.}
\label{tab:golden}
\resizebox{\textwidth}{!}{
\begin{tabular}{llcccccccccccc}
\toprule
\multirow{3}{*}{\textbf{Benchmark}} & \multirow{3}{*}{\textbf{Finetuning Dataset}}
& \multicolumn{6}{c}{\cellcolor{dustyblue}\textbf{Open-Sourced LLMs}}
& \multicolumn{6}{c}{\cellcolor{sagegreen}\textbf{Closed-Sourced LLMs}} \\
& & \multicolumn{3}{c}{\cellcolor{dustyblue}\textbf{gpt-oss-120b}}
    & \multicolumn{3}{c}{\cellcolor{dustyblue}\textbf{Qwen3-14B}}
    & \multicolumn{3}{c}{\cellcolor{sagegreen}\textbf{Gemini-2.5-Pro}}
    & \multicolumn{3}{c}{\cellcolor{sagegreen}\textbf{GPT-5}} \\
\cmidrule(lr){3-5} \cmidrule(lr){6-8} \cmidrule(lr){9-11} \cmidrule(lr){12-14}
& 
& \textbf{P} & \textbf{R} & \textbf{F1}
& \textbf{P} & \textbf{R} & \textbf{F1}
& \textbf{P} & \textbf{R} & \textbf{F1}
& \textbf{P} & \textbf{R} & \textbf{F1} \\
\midrule
HumanEval     & Evol-CodeAlpaca & 0.75 & 0.80 & 0.78 & 0.83 & 0.85 & 0.84 & 0.83 & 0.88 & 0.85 & 0.87 & 0.91 & 0.90 \\
HumanEval     & CodeAlpaca      & 0.90 & 0.90 & 0.90 & 0.79 & 0.82 & 0.80 & 0.88 & 0.92 & 0.90 & 0.98 & 0.99 & 0.99 \\
MBPP          & CodeAlpaca      & 0.81 & 0.83 & 0.82 & 0.71 & 0.72 & 0.72 & 0.84 & 0.84 & 0.84 & 0.91 & 0.97 & 0.94 \\
HumanEval     & Magicoder       & 0.80 & 0.86 & 0.83 & 0.85 & 0.86 & 0.86 & 0.91 & 0.90 & 0.90 & 0.95 & 0.92 & 0.93 \\
LiveCodeBench & CodeAlpaca      & 0.62 & 0.65 & 0.63 & 0.66 & 0.66 & 0.66 & 0.90 & 0.93 & 0.92 & 0.86 & 0.94 & 0.90 \\
MBPP          & Magicoder       & 0.70 & 0.76 & 0.73 & 0.82 & 0.88 & 0.85 & 0.65 & 0.70 & 0.68 & 0.87 & 0.80 & 0.83 \\
LiveCodeBench & Evol-CodeAlpaca & 0.76 & 0.75 & 0.75 & 0.53 & 0.56 & 0.54 & 0.83 & 0.81 & 0.82 & 0.90 & 0.91 & 0.90 \\
LiveCodeBench & Magicoder       & 0.84 & 0.81 & 0.83 & 0.73 & 0.77 & 0.75 & 0.72 & 0.77 & 0.75 & 0.94 & 0.90 & 0.92 \\
MBPP          & Evol-CodeAlpaca & 0.67 & 0.66 & 0.67 & 0.50 & 0.55 & 0.52 & 0.63 & 0.64 & 0.64 & 0.81 & 0.73 & 0.77 \\
\midrule
\textbf{Overall} & --
& \textbf{0.78} & \textbf{0.78} & \textbf{0.78}
& \textbf{0.71} & \textbf{0.73} & \textbf{0.72}
& \textbf{0.81} & \textbf{0.82} & \textbf{0.81}
& \textbf{0.91} & \textbf{0.90} & \textbf{0.91} \\
\bottomrule
\end{tabular}
}
\end{table*}

Table~\ref{tab:golden} reports overall fine-grained results across the nine benchmark-training combinations, and Table~\ref{tab:golden_type} further breaks down results by contamination category. Overall, \tool achieves strong performance across all evaluated LLM backbones, with macro-average F1 ranging from 0.72 to 0.91. 

The benchmark-training breakdown further indicates that contamination detection difficulty varies across settings. Combinations involving \textsc{HumanEval} with \textsc{CodeAlpaca} or \textsc{Magicoder} are consistently among the easiest across LLM backbones, whereas \textsc{MBPP}--\textsc{Evol-CodeAlpaca} and \textsc{LiveCodeBench}--\textsc{Evol-CodeAlpaca} are consistently more challenging. This pattern may reflect differences in how benchmark style interacts with the characteristics of the post-training corpus. \textsc{HumanEval} consists of relatively canonical function-synthesis tasks with compact specifications, which may make contamination cases easier to recognize when paired with post-training corpora. In contrast, \textsc{MBPP} contains shorter and often more diverse task descriptions, while \textsc{LiveCodeBench} introduces newer and stylistically broader problems. When these are paired with \textsc{Evol-CodeAlpaca}, whose instruction evolution process tends to introduce greater paraphrastic and structural variation, the boundary between contamination categories becomes harder to determine. 

The category-level breakdown also reveals a clear and consistent pattern across LLM backbones. Performance is strongest on the {\em Unrelated} and {\em Functionally Identical} categories and remains robust on the {\em Functionally Identical} and {\em Shared Logic} categories when using strong LLMs such as GPT-5. This indicates that \tool is effective both at filtering out clearly unrelated pairs and at identifying semantically meaningful overlap beyond near-duplicate matching. By contrast, with weaker models, \tool faces challenges in distinguishing {\em Nearly Identical} pairs from {\em Shared Logic} pairs, where the model must determine the degree of semantic differences between two similar tasks. 

\begin{table*}[t]
\centering
\small
\caption{Per-category fine-grained contamination detection results on the evaluation benchmark.}
\label{tab:golden_type}
\resizebox{\textwidth}{!}{
\begin{tabular}{lcccccccccccc}
\toprule
\multirow{2}{*}{\textbf{Model}}
& \multicolumn{3}{c}{\textbf{Functionally Identical}}
& \multicolumn{3}{c}{\textbf{Nearly Identical}}
& \multicolumn{3}{c}{\textbf{Shared Logic}}
& \multicolumn{3}{c}{\textbf{Unrelated}} \\
\cmidrule(lr){2-4} \cmidrule(lr){5-7} \cmidrule(lr){8-10} \cmidrule(lr){11-13}
& \textbf{P} & \textbf{R} & \textbf{F1}
& \textbf{P} & \textbf{R} & \textbf{F1}
& \textbf{P} & \textbf{R} & \textbf{F1}
& \textbf{P} & \textbf{R} & \textbf{F1} \\
\midrule
\headerblue \multicolumn{13}{c}{\textit{Open-Sourced LLMs}} \\
\midrule
gpt-oss-120b & 0.79 & 0.88 & 0.83 & 0.85 & 0.64 & 0.73 & 0.60 & 0.72 & 0.66 & 0.88 & 0.88 & 0.88 \\
Qwen3-14B    & 0.93 & 1.00 & \textbf{0.96} & 0.71 & 0.52 & 0.60 & 0.60 & 0.55 & 0.57 & 0.60 & 0.80 & 0.69 \\
\midrule
\headergreen \multicolumn{13}{c}{\textit{Closed-Sourced LLMs}} \\
\midrule
Gemini-2.5-Pro & 0.89 & 0.99 & 0.94 & 0.69 & 0.84 & 0.76 & 0.74 & 0.50 & 0.60 & 0.92 & 0.94 & 0.93 \\
GPT-5          & 1.00 & 0.93 & \textbf{0.96} & 0.82 & 0.88 & \textbf{0.85} & 0.84 & 0.84 & \textbf{0.84} & 0.95 & 0.95 & \textbf{0.95} \\
\bottomrule
\end{tabular}
}
\end{table*}

\paragraph{Binary Contamination Detection.}

\begin{table}[t]
\centering
\small
\caption{Comparison with binary detection methods.}
\label{tab:main_baseline}
\begin{tabular}{lccc}
\toprule
\textbf{Method} & \textbf{P} & \textbf{R} & \textbf{F1} \\
\midrule
\textsc{ROOTS}~\citep{piktus2023roots} & 0.57 & 0.58 & 0.57 \\
\textsc{PSM}~\citep{riddell2024quantifying} & 0.76 & 0.57 & 0.65 \\
\textsc{LLM Decontaminator}~\citep{yang2023rethinking} & \textbf{1.00} & 0.17 & 0.29 \\
\midrule
\tool & 0.93 & \textbf{0.91} & \textbf{0.92} \\
\bottomrule
\end{tabular}
\end{table}

We next evaluate a binary reduction of the task only for comparison with existing methods. We use GPT-5 as the backbone for both \tool and \textsc{LLM Decontaminator}. Table~\ref{tab:main_baseline} shows that \tool substantially outperforms all three baselines, achieving an F1 score of 0.92. 
In particular, we observed that  existing methods miss many contaminated benchmark--training pairs with non-trivial semantic overlap, leading to a low recall (0.17-0.58).

\paragraph{Ablation Study.}
\label{sec:ablation}

We further evaluate the contribution of each stage in \tool through four targeted ablations with GPT-5 as the LLM backbone. 

To assess the role of instruction normalization, we remove the task description normalization stage and perform embedding-based triage and LLM-based verification directly on the original benchmark and post-training task descriptions. Without normalization, performance in the fine-grained setting drops from a precision of 0.91, a recall of 0.90, and an F1 score of 0.91 to a precision of 0.73, a recall of 0.72, and an F1 score of 0.72. This result suggests that normalization improves contamination detection by reducing superficial variation in formatting and phrasing, thereby making semantically related task pairs more comparable before downstream triage and verification.

To examine the role of semantic triage, we remove the embedding-based stage and directly pass all benchmark-training task pairs to LLM-based verification. Removing triage changes predictive performance only slightly, from a precision of 0.91, a recall of 0.90, and an F1 score of 0.91 to a precision of 0.91, a recall of 0.87, and an F1 score of 0.89. However, it significantly increases the number of pairs passed to LLM verification from 13,489 to 171,654, raising the total token cost by 1172\%. This shows that embedding-based triage greatly improves efficiency and scalability by narrowing the verification space, while having limited influence on classification accuracy.

To examine the role of LLM-based fine-grained verification, we remove the verification stage and replace it with threshold-only classifiers based solely on embedding similarity. We evaluate this ablation in both the fine-grained and binary settings. For the fine-grained setting, we introduce three similarity thresholds and tune them jointly on the development split by maximizing macro-F1. Pairs with similarity scores above $t_1$ are classified as Functionally Identical, pairs between $t_1$ and $t_2$ as Nearly Identical, pairs between $t_2$ and $t_3$ as Shared Logic, and pairs below $t_3$ as Unrelated. Appendix~\ref{sec:threshold} visualizes the macro-F1 landscape over the three-threshold search space. On the held-out test split, this fine-grained threshold-only variant achieves a macro-average precision of 0.39, recall of 0.38, and F1 of 0.37, substantially below the full \tool pipeline with GPT-5, which achieves 0.91 precision, 0.90 recall, and 0.91 F1. We also evaluate a binary threshold-only variant by collapsing Functionally Identical, Nearly Identical, and Shared Logic into contaminated and tuning a single decision threshold on the development split. This binary variant achieves an F1 score of 0.58, compared with 0.92 for full \tool in the same binary setting. Appendix~\ref{sec:threshold} reports the corresponding binary threshold sensitivity curve. The clear degradation in both settings indicates that coarse similarity signals alone are insufficient for reliable contamination detection and that explicit verification via LLM reasoning over task intent and solution strategy is necessary for accurate categorization.

Finally, we remove the trivial-task filtering stage and retain all pairs predicted as contaminated after verification. Without this stage, precision decreases from 0.91 to 0.75, recall decreases from 0.90 to 0.74, and F1 decreased from 0.91 to 0.74. In addition, trivial-task filtering removes 18.61\% of predicted contaminated pairs from the final output. This result shows that the final filtering stage mainly improves the specificity of the reported contamination set by excluding generic utility-style problems that would otherwise inflate contamination estimates.

\begin{table}[t]
\centering
\small
\caption{Token usage and cost using \textsc{GPT-5} with and without embedding-based triage.}
\label{tab:token_cost}
\begin{tabular}{lcc}
\toprule
\textbf{Setting} & \textbf{Tokens} & \textbf{Cost} \\
\midrule
\tool{} (no triage)   & 163.59M & \$1,635.9 \\
\tool{} (with triage) & 12.86M  & \$128.6 \\
\bottomrule
\end{tabular}
\end{table}

\paragraph{Token Cost Analysis.} Table~\ref{tab:token_cost} shows the effect of embedding-based triage on LLM verification cost. By filtering candidate pairs before fine-grained verification, \tool reduces total token usage from 163.59 M to 12.86 M, a reduction of 92.1\%. Under the same pricing setting, the estimated cost decreases from \$1,635.9 to \$128.6 for GPT-5.

\subsection{Failure Case Analysis}
\label{sec:failure-case}

To better understand the limitations of \tool, we manually inspect 59 misclassified pairs produced by \textsc{GPT-5} on the benchmark. The dominant error source is \emph{core logic misidentification} (50.85\%), where the model fails to abstract the shared algorithmic idea behind two tasks and instead over-relies on surface framing or local constraints. A second major source is \emph{confusion between adjacent semantic categories} (37.29\%), especially between Nearly Identical and Shared Logic, reflecting the difficulty of making fine-grained boundary-sensitive judgments. Less frequent errors arise from hallucinated or internally inconsistent reasoning (8.47\%), as well as residual formatting artifacts that survive normalization (3.39\%). These error patterns suggest that future improvements will require more reliable abstraction of task-level computational logic and sharper discrimination between neighboring contamination categories. Detailed breakdowns and examples are provided in Appendix~\ref{sec:failure-cases}.

\section{Conclusion}

We presented \tool, a semantic-aware multi-stage framework for fine-grained code contamination detection in Code LLM evaluation. By formulating contamination detection as a fine-grained semantic categorization problem, \tool goes beyond prior binary detection methods and enables a more precise analysis of coding task overlap. We also constructed the first benchmark for fine-grained code contamination detection. Experiments show that \tool achieves strong performance across multiple LLM backbones and substantially outperforms existing methods in binary comparison. Overall, our results highlight the importance of combining semantic retrieval, LLM-based verification, and specificity-aware filtering for reliable code contamination detection.

\section{Limitations}
\label{sec:limitation}

Our benchmark is constructed by sampling from similarity-based candidate pools rather than exhaustively annotating the full benchmark--training cross-product, which may underrepresent rare or highly indirect contamination patterns. In addition, \tool depends on both embedding-based triage and LLM-based verification: relevant pairs may be pruned before verification, and the LLM backbone can still make boundary-sensitive errors on closely related categories. Our current pipeline also operates primarily on task descriptions, with reference solutions used only during manual annotation, and may therefore miss signals available from richer program-level evidence. Finally, although triage substantially improves efficiency, fine-grained verification still incurs non-trivial LLM inference cost, which may limit applicability in very large-scale or latency-sensitive settings.

\bibliographystyle{plainnat}
\bibliography{custom}

@inproceedings{chen2021evaluating,
  title={Evaluating Large Language Models Trained on Code},
  author={Chen, Mark and Tworek, Jerry and Jun, Heewoo and Yuan, Qiming and Ponde de Oliveira Pinto, Henrique and Kaplan, Jared and Edwards, Harri and Burda, Yuri and others},
  booktitle={Advances in Neural Information Processing Systems (NeurIPS)},
  year={2021}
}

@inproceedings{austin2021program,
  title={Program synthesis with large language models},
  author={Austin, Jacob and Odena, Augustus and Nye, Maxwell and Bosma, Maarten and Michalewski, Henryk and Dohan, David and Jiang, Ellen and Cai, Carrie and Terry, Michael and Le, Quoc and Sutton, Charles},
  booktitle={Advances in Neural Information Processing Systems (NeurIPS)},
  volume={34},
  pages={ 17981--17993 },
  year={2021}
}

@article{yang2023rethinking,
  title={Rethinking Benchmark and Contamination for Language Models with Rephrased Samples},
  author={Yang, Shuo and Chiang, Wei-Lin and Zheng, Lianmin and Gonzalez, Joseph E. and Stoica, Ion},
  journal={arXiv preprint arXiv:2311.04850},
  year={2023}
}

@inproceedings{riddell2024quantifying,
  title={Quantifying contamination in evaluating code generation capabilities of language models},
  author={Riddell, Martin and Ni, Ansong and Cohan, Arman},
  booktitle={Proceedings of the 62nd Annual Meeting of the Association for Computational Linguistics (Volume 1: Long Papers)},
  pages={14116--14137},
  year={2024}
}

@article{roy2007survey,
  title={A survey on software clone detection research},
  author={Roy, Chanchal Kumar and Cordy, James R},
  journal={Queen’s School of computing TR},
  volume={541},
  number={115},
  pages={64--68},
  year={2007}
}

@inproceedings{jain2025livecodebench,
  title={{LiveCodeBench}: Holistic and Contamination-Free Evaluation of Large Language Models for Code},
  author={Jain, Naman and Han, King and Gu, Alex and Li, Wen-Ding and Yan, Fanjia and Zhang, Tianjun and Wang, Sida and Solar-Lezama, Armando and Sen, Koushik and Stoica, Ion},
  booktitle={International Conference on Learning Representations (ICLR)},
  year={2025}
}

@misc{codealpaca,
  author = {Sahil Chaudhary},
  title = {Code Alpaca: An Instruction-following LLaMA model for code generation},
  year = {2023},
  publisher = {GitHub},
  journal = {GitHub repository},
  howpublished = {\url{https://github.com/sahil280114/codealpaca}},
}

@article{luo2023wizardcoder,
  title={Wizardcoder: Empowering code large language models with evol-instruct},
  author={Luo, Ziyang and Xu, Can and Zhao, Pu and Sun, Qingfeng and Geng, Xiubo and Hu, Wenxiang and Tao, Chongyang and Ma, Jing and Lin, Qingwei and Jiang, Daxin},
  journal={arXiv preprint arXiv:2306.08568},
  year={2023}
}

@article{wei2023magicoder,
  title={Magicoder: Empowering code generation with oss-instruct},
  author={Wei, Yuxiang and Wang, Zhe and Liu, Jiawei and Ding, Yifeng and Zhang, Lingming},
  journal={arXiv preprint arXiv:2312.02120},
  year={2023}
}

@article{xu2024benchmark,
  title={Benchmark data contamination of large language models: A survey},
  author={Xu, Cheng and Guan, Shuhao and Greene, Derek and Kechadi, M and others},
  journal={arXiv preprint arXiv:2406.04244},
  year={2024}
}

@inproceedings{dong2024generalization,
  title={Generalization or memorization: Data contamination and trustworthy evaluation for large language models},
  author={Dong, Yihong and Jiang, Xue and Liu, Huanyu and Jin, Zhi and Gu, Bin and Yang, Mengfei and Li, Ge},
  booktitle={Findings of the Association for Computational Linguistics: ACL 2024},
  pages={12039--12050},
  year={2024}
}

@inproceedings{balloccu2024leak,
  title={Leak, cheat, repeat: Data contamination and evaluation malpractices in closed-source LLMs},
  author={Balloccu, Simone and Schmidtov{\'a}, Patr{\'\i}cia and Lango, Mateusz and Du{\v{s}}ek, Ond{\v{r}}ej},
  booktitle={Proceedings of the 18th Conference of the European Chapter of the Association for Computational Linguistics (Volume 1: Long Papers)},
  pages={67--93},
  year={2024}
}

@inproceedings{deng2024investigating,
  title={Investigating data contamination in modern benchmarks for large language models},
  author={Deng, Chunyuan and Zhao, Yilun and Tang, Xiangru and Gerstein, Mark and Cohan, Arman},
  booktitle={Proceedings of the 2024 Conference of the North American Chapter of the Association for Computational Linguistics: Human Language Technologies (Volume 1: Long Papers)},
  pages={8706--8719},
  year={2024}
}

@article{brown2020language,
  title={Language models are few-shot learners},
  author={Brown, Tom and Mann, Benjamin and Ryder, Nick and Subbiah, Melanie and Kaplan, Jared D and Dhariwal, Prafulla and Neelakantan, Arvind and Shyam, Pranav and Sastry, Girish and Askell, Amanda and others},
  journal={Advances in neural information processing systems},
  volume={33},
  pages={1877--1901},
  year={2020}
}

@article{chowdhery2023palm,
  title={Palm: Scaling language modeling with pathways},
  author={Chowdhery, Aakanksha and Narang, Sharan and Devlin, Jacob and Bosma, Maarten and Mishra, Gaurav and Roberts, Adam and Barham, Paul and Chung, Hyung Won and Sutton, Charles and Gehrmann, Sebastian and others},
  journal={Journal of machine learning research},
  volume={24},
  number={240},
  pages={1--113},
  year={2023}
}

@article{achiam2023gpt,
  title={Gpt-4 technical report},
  author={Achiam, Josh and Adler, Steven and Agarwal, Sandhini and Ahmad, Lama and Akkaya, Ilge and Aleman, Florencia Leoni and Almeida, Diogo and Altenschmidt, Janko and Altman, Sam and Anadkat, Shyamal and others},
  journal={arXiv preprint arXiv:2303.08774},
  year={2023}
}

@article{touvron2023llama,
  title={Llama 2: Open foundation and fine-tuned chat models},
  author={Touvron, Hugo and Martin, Louis and Stone, Kevin and Albert, Peter and Almahairi, Amjad and Babaei, Yasmine and Bashlykov, Nikolay and Batra, Soumya and Bhargava, Prajjwal and Bhosale, Shruti and others},
  journal={arXiv preprint arXiv:2307.09288},
  year={2023}
}

@inproceedings{piktus2023roots,
  title={The ROOTS search tool: Data transparency for LLMs},
  author={Piktus, Aleksandra and Akiki, Christopher and Villegas, Paulo and Lauren{\c{c}}on, Hugo and Dupont, G{\'e}rard and Luccioni, Sasha and Jernite, Yacine and Rogers, Anna},
  booktitle={Proceedings of the 61st Annual Meeting of the Association for Computational Linguistics (Volume 3: System Demonstrations)},
  pages={304--314},
  year={2023}
}

@inproceedings{witteveen2019paraphrasing,
  title={Paraphrasing with large language models},
  author={Witteveen, Sam and Andrews, Martin},
  booktitle={Proceedings of the 3rd Workshop on Neural Generation and Translation},
  pages={215--220},
  year={2019}
}

@inproceedings{abaskohi2023lm,
  title={LM-CPPF: Paraphrasing-guided data augmentation for contrastive prompt-based few-shot fine-tuning},
  author={Abaskohi, Amirhossein and Rothe, Sascha and Yaghoobzadeh, Yadollah},
  booktitle={Proceedings of the 61st Annual Meeting of the Association for Computational Linguistics (Volume 2: Short Papers)},
  pages={670--681},
  year={2023}
}

@article{lee2023platypus,
  title={Platypus: Quick, cheap, and powerful refinement of llms},
  author={Lee, Ariel N and Hunter, Cole J and Ruiz, Nataniel},
  journal={arXiv preprint arXiv:2308.07317},
  year={2023}
}

@article{gunasekar2023textbooks,
  title={Textbooks are all you need},
  author={Gunasekar, Suriya and Zhang, Yi and Aneja, Jyoti and Mendes, Caio C{\'e}sar Teodoro and Del Giorno, Allie and Gopi, Sivakanth and Javaheripi, Mojan and Kauffmann, Piero and de Rosa, Gustavo and Saarikivi, Olli and others},
  journal={arXiv preprint arXiv:2306.11644},
  year={2023}
}

@article{ain2019systematic,
  title={A systematic review on code clone detection},
  author={Ain, Qurat Ul and Butt, Wasi Haider and Anwar, Muhammad Waseem and Azam, Farooque and Maqbool, Bilal},
  journal={IEEE access},
  volume={7},
  pages={86121--86144},
  year={2019},
  publisher={IEEE}
}

@inproceedings{sajnani2016sourcerercc,
  title={Sourcerercc: Scaling code clone detection to big-code},
  author={Sajnani, Hitesh and Saini, Vaibhav and Svajlenko, Jeffrey and Roy, Chanchal K and Lopes, Cristina V},
  booktitle={Proceedings of the 38th international conference on software engineering},
  pages={1157--1168},
  year={2016}
}

@inproceedings{svajlenko2015evaluating,
  title={Evaluating clone detection tools with bigclonebench},
  author={Svajlenko, Jeffrey and Roy, Chanchal K},
  booktitle={2015 IEEE international conference on software maintenance and evolution (ICSME)},
  pages={131--140},
  year={2015},
  organization={IEEE}
}

@article{sturua2024jina,
  title={jina-embeddings-v3: Multilingual embeddings with task lora},
  author={Sturua, Saba and Mohr, Isabelle and Akram, Mohammad Kalim and G{\"u}nther, Michael and Wang, Bo and Krimmel, Markus and Wang, Feng and Mastrapas, Georgios and Koukounas, Andreas and Wang, Nan and others},
  journal={arXiv preprint arXiv:2409.10173},
  year={2024}
}

@inproceedings{lin2021pyserini,
  title={Pyserini: A Python toolkit for reproducible information retrieval research with sparse and dense representations},
  author={Lin, Jimmy and Ma, Xueguang and Lin, Sheng-Chieh and Yang, Jheng-Hong and Pradeep, Ronak and Nogueira, Rodrigo},
  booktitle={Proceedings of the 44th international ACM SIGIR conference on research and development in information retrieval},
  pages={2356--2362},
  year={2021}
}

@article{cheng2025survey,
  title={A survey on data contamination for large language models},
  author={Cheng, Yuxing and Chang, Yi and Wu, Yuan},
  journal={arXiv preprint arXiv:2502.14425},
  year={2025}
}

@article{ravaut2024comprehensive,
  title={A comprehensive survey of contamination detection methods in large language models},
  author={Ravaut, Mathieu and Ding, Bosheng and Jiao, Fangkai and Chen, Hailin and Li, Xingxuan and Zhao, Ruochen and Qin, Chengwei and Xiong, Caiming and Joty, Shafiq},
  journal={arXiv preprint arXiv:2404.00699},
  year={2024}
}

@article{dekoninck2024evading,
  title={Evading data contamination detection for language models is (too) easy},
  author={Dekoninck, Jasper and M{\"u}ller, Mark Niklas and Baader, Maximilian and Fischer, Marc and Vechev, Martin},
  journal={arXiv preprint arXiv:2402.02823},
  year={2024}
}

@article{zhao2024cap,
  title={Cap: Data contamination detection via consistency amplification},
  author={Zhao, Yi and Li, Jing and Yang, Linyi},
  journal={arXiv preprint arXiv:2410.15005},
  year={2024}
}

@article{duan2024membership,
  title={Do membership inference attacks work on large language models?},
  author={Duan, Michael and Suri, Anshuman and Mireshghallah, Niloofar and Min, Sewon and Shi, Weijia and Zettlemoyer, Luke and Tsvetkov, Yulia and Choi, Yejin and Evans, David and Hajishirzi, Hannaneh},
  journal={arXiv preprint arXiv:2402.07841},
  year={2024}
}

@inproceedings{fu2025does,
  title={Does data contamination detection work (well) for llms? a survey and evaluation on detection assumptions},
  author={Fu, Yujuan and Uzuner, Ozlem and Yetisgen-Yildiz, Meliha and Xia, Fei},
  booktitle={Findings of the Association for Computational Linguistics: NAACL 2025},
  pages={5235--5256},
  year={2025}
}

@article{shi2023detecting,
  title={Detecting pretraining data from large language models},
  author={Shi, Weijia and Ajith, Anirudh and Xia, Mengzhou and Huang, Yangsibo and Liu, Daogao and Blevins, Terra and Chen, Danqi and Zettlemoyer, Luke},
  journal={arXiv preprint arXiv:2310.16789},
  year={2023}
}

@article{zhang2024min,
  title={Min-k\%++: Improved baseline for detecting pre-training data from large language models},
  author={Zhang, Jingyang and Sun, Jingwei and Yeats, Eric and Ouyang, Yang and Kuo, Martin and Zhang, Jianyi and Yang, Hao Frank and Li, Hai},
  journal={arXiv preprint arXiv:2404.02936},
  year={2024}
}

@article{rajore2024truce,
  title={Truce: Private benchmarking to prevent contamination and improve comparative evaluation of llms},
  author={Rajore, Tanmay and Chandran, Nishanth and Sitaram, Sunayana and Gupta, Divya and Sharma, Rahul and Mittal, Kashish and Swaminathan, Manohar},
  journal={arXiv preprint arXiv:2403.00393},
  year={2024}
}

@inproceedings{jacovi2023stop,
  title={Stop uploading test data in plain text: Practical strategies for mitigating data contamination by evaluation benchmarks},
  author={Jacovi, Alon and Caciularu, Avi and Goldman, Omer and Goldberg, Yoav},
  booktitle={Proceedings of the 2023 Conference on Empirical Methods in Natural Language Processing},
  pages={5075--5084},
  year={2023}
}

@inproceedings{li2024latesteval,
  title={Latesteval: Addressing data contamination in language model evaluation through dynamic and time-sensitive test construction},
  author={Li, Yucheng and Guerin, Frank and Lin, Chenghua},
  booktitle={Proceedings of the AAAI Conference on Artificial Intelligence},
  volume={38},
  number={17},
  pages={18600--18607},
  year={2024}
}

@article{white2024livebench,
  title={Livebench: A challenging, contamination-free llm benchmark},
  author={White, Colin and Dooley, Samuel and Roberts, Manley and Pal, Arka and Feuer, Ben and Jain, Siddhartha and Shwartz-Ziv, Ravid and Jain, Neel and Saifullah, Khalid and Naidu, Siddartha and others},
  journal={arXiv preprint arXiv:2406.19314},
  volume={4},
  pages={2},
  year={2024}
}

@inproceedings{zhu2024clean,
  title={CLEAN--EVAL: Clean evaluation on contaminated large language models},
  author={Zhu, Wenhong and Hao, Hongkun and He, Zhiwei and Song, Yun-Ze and Yueyang, Jiao and Zhang, Yumeng and Hu, Hanxu and Wei, Yiran and Wang, Rui and Lu, Hongyuan},
  booktitle={Findings of the Association for Computational Linguistics: NAACL 2024},
  pages={835--847},
  year={2024}
}

@inproceedings{li2023cleva,
  title={Cleva: Chinese language models evaluation platform},
  author={Li, Yanyang and Zhao, Jianqiao and Zheng, Duo and Hu, Zi-Yuan and Chen, Zhi and Su, Xiaohui and Huang, Yongfeng and Huang, Shijia and Lin, Dahua and Lyu, Michael R and others},
  booktitle={Proceedings of the 2023 Conference on Empirical Methods in Natural Language Processing: System Demonstrations},
  pages={186--217},
  year={2023}
}

@article{zakeri2023systematic,
  title={A systematic literature review on source code similarity measurement and clone detection: Techniques, applications, and challenges},
  author={Zakeri-Nasrabadi, Morteza and Parsa, Saeed and Ramezani, Mohammad and Roy, Chanchal and Ekhtiarzadeh, Masoud},
  journal={Journal of Systems and Software},
  volume={204},
  pages={111796},
  year={2023},
  publisher={Elsevier}
}

@inproceedings{li2017cclearner,
  title={Cclearner: A deep learning-based clone detection approach},
  author={Li, Liuqing and Feng, He and Zhuang, Wenjie and Meng, Na and Ryder, Barbara},
  booktitle={2017 IEEE international conference on software maintenance and evolution (ICSME)},
  pages={249--260},
  year={2017},
  organization={IEEE}
}

@inproceedings{johnson1994substring,
  title={Substring Matching for Clone Detection and Change Tracking.},
  author={Johnson, J Howard},
  booktitle={ICSM},
  volume={94},
  pages={120--126},
  year={1994}
}

@article{kamiya2002ccfinder,
  title={CCFinder: A multilinguistic token-based code clone detection system for large scale source code},
  author={Kamiya, Toshihiro and Kusumoto, Shinji and Inoue, Katsuro},
  journal={IEEE transactions on software engineering},
  volume={28},
  number={7},
  pages={654--670},
  year={2002},
  publisher={IEEE}
}

@inproceedings{roy2008nicad,
  title={NICAD: Accurate detection of near-miss intentional clones using flexible pretty-printing and code normalization},
  author={Roy, Chanchal K and Cordy, James R},
  booktitle={2008 16th iEEE international conference on program comprehension},
  pages={172--181},
  year={2008},
  organization={IEEE}
}

@inproceedings{jiang2007deckard,
  title={Deckard: Scalable and accurate tree-based detection of code clones},
  author={Jiang, Lingxiao and Misherghi, Ghassan and Su, Zhendong and Glondu, Stephane},
  booktitle={29th International Conference on Software Engineering (ICSE'07)},
  pages={96--105},
  year={2007},
  organization={IEEE}
}

@inproceedings{krinke2001identifying,
  title={Identifying similar code with program dependence graphs},
  author={Krinke, Jens},
  booktitle={Proceedings eighth working conference on reverse engineering},
  pages={301--309},
  year={2001},
  organization={IEEE}
}

@inproceedings{gabel2008scalable,
  title={Scalable detection of semantic clones},
  author={Gabel, Mark and Jiang, Lingxiao and Su, Zhendong},
  booktitle={Proceedings of the 30th international conference on Software engineering},
  pages={321--330},
  year={2008}
}

@inproceedings{white2016deep,
  title={Deep learning code fragments for code clone detection},
  author={White, Martin and Tufano, Michele and Vendome, Christopher and Poshyvanyk, Denys},
  booktitle={Proceedings of the 31st IEEE/ACM international conference on automated software engineering},
  pages={87--98},
  year={2016}
}

@inproceedings{zhang2017automated,
  title={Automated transplantation and differential testing for clones},
  author={Zhang, Tianyi and Kim, Miryung},
  booktitle={2017 IEEE/ACM 39th International Conference on Software Engineering (ICSE)},
  pages={665--676},
  year={2017},
  organization={IEEE}
}


\appendix

\section{Prompts in \tool}
\label{sec:prompts}

This section presents all the prompts of \tool. Table~\ref{tab:normalization-prompt} shows the prompt of the instruction normalization stage described in Section~\ref{sec:normalization}. Table~\ref{tab:verification-prompt} shows the prompt of the fine-grained verification stage (Section~\ref{sec:fine-grained-verification}). Table~\ref{tab:final-screening-prompt} shows the prompt of the final screening stage (Section~\ref{sec:final-screening}).

\begin{table}[ht]
    \centering
    \small
    \caption{Prompt for the instruction normalization stage.}
    \label{tab:normalization-prompt}
    \begin{tabular}{>{\raggedright\arraybackslash\ttfamily}p{0.98\linewidth}}
        \toprule
            \headercolorlong
            \textbf{Instruction}\\
            Carefully read the programming task and the examples provided. Then rephrase the original task description into clean and concise ones. Make sure the rephrased task description follow the style and length of rephrased ones provided in the examples. Directly return the rephrased task description.\\\\

            \headercolorlong
            \textbf{Example 1}\\
            \textbf{Original Task Description}\\
            You are tasked with implementing a function to convert an image wrapper to a OpenGL texture. The image wrapper is a data structure that holds image data, and the OpenGL texture is a representation of the image suitable for rendering in an OpenGL environment.\\

            \textbf{Rephrased Task Description}\\
            Implement a method that converts an image wrapper containing image data into a texture suitable for rendering in an OpenGL environment. The method should handle data formatting and texture creation within a properly initialized OpenGL context, ensuring correctness and efficiency.\\\\

            \headercolorlong
            \textbf{Example 2}\\
            \textbf{Original Task Description}\\
            Write a function to find the maximum value in record list as tuple attribute in the given tuple list.\\
            \textbf{Rephrased Task Description}\\
            Implement a method that processes a list of tuples and returns the maximum value found among a specific attribute within each tuple. The method should correctly extract and compare values to determine the highest one.\\\\

            \headercolorlong
            \textbf{Task}\\
            \textbf{Original Task Description}\\
            \{original\_description\}\\
            \textbf{Rephrased Task Description}\\
            {}[New description here]\\
        \bottomrule
    \end{tabular}
\end{table}

\begin{table*}[p]
    \centering
    \small
    \caption{Prompt for the fine-grained verification stage.}
    \label{tab:verification-prompt}
    \begin{adjustbox}{max width=\textwidth, max totalheight=\textheight, keepaspectratio}
    \begin{tabular}{>{\raggedright\arraybackslash}p{0.98\textwidth}}
        \toprule
        \begin{minipage}[t]{0.485\textwidth}
            \ttfamily
            \promptheader{Instruction}\\
            1. You will see two tasks: Task A and Task B.\\
            2. Read both carefully, noting their goals, inputs/outputs, and logic.\\
            3. Choose the single most accurate relationship from the categories below.\\\\

            \promptheader{Relationship Categories}\\
            A. Functionally Identical\\
            Choose this if the tasks are perfect duplicates. They accomplish the exact same goal, take the same kinds of input, and produce the same kinds of output. They are essentially two descriptions of the very same problem.\\
            Litmus Test: Could the solution for one task solve the other with zero changes? If yes, choose A. Otherwise, do NOT choose A.\\\\

            B. Nearly Identical (Variation of the Same Problem)\\
            Choose this if the tasks solve the same fundamental problem, but differ only in minor surface details, but share all the same core logic. They solve the same fundamental problem, but with minor differences in constraints, data types, or input/output formats.\\
            Litmus Test: If the tasks are not perfectly identical (A fails), but the **core logic is identical**, choose B. If the core logic differs, do NOT choose B.\\\\

            C. Shared Logic (Different Problems, Same Algorithm)\\
            Choose this if the tasks solve different problems using the same algorithmic method. The tasks have different goals and may come from unrelated domains, but they are solved using the same core algorithm or logical procedure.\\
            Litmus Test: If neither A nor B applies, but the algorithmic approach is the same, choose C. If the algorithm differs, do NOT choose C.\\\\

            D. Unrelated or Different Domain\\
            Choose this if the tasks do not share the same algorithmic logic. This includes two cases:\\
            The tasks are from the same general domain (e.g., both deal with arrays or graphs) but require different algorithms or solution methods.\\
            The tasks are completely unrelated --- they have no meaningful conceptual, logical, or domain connection.\\
            Litmus Test: If none of A, B, or C applies, choose D.\\
        \end{minipage}
        \hfill
        \begin{minipage}[t]{0.485\textwidth}
            \ttfamily
            \promptheader{Examples}\\
            \textbf{Example 1}\\
            Task A:\\
            Determine if a given string is a palindrome, returning True if it reads the same backward as forward.\\
            Task B:\\
            Implement a method in Ruby that determines whether a given string is a palindrome.\\
            Answer: A\\\\

            \textbf{Example 2}\\
            Task A:\\
            Generate a space-delimited string of numbers starting from 0 up to n inclusive.\\
            Task B:\\
            Implement a C++ function to print the numbers from 0 to n in ascending order.\\
            Answer: B\\\\

           \textbf{Example 3}\\
            Task A:\\
            Given an array of integers and a positive integer k, return a sorted list of the k largest numbers in the array.\\
            Task B:\\
            Implement a function to identify the two largest numbers in an array and return them in descending order.\\
            Answer: C\\\\

            \textbf{Example 4}\\
            Task A:\\
            Determine if any two numbers in the given list are closer to each other than a specified threshold.\\
            Task B:\\
            Given a sorted integer array and two integers k and x, return the k closest integers to x, sorted in ascending order. An integer is considered closer to x if it has a smaller absolute difference, or the same difference but is smaller in value.\\
            Answer: D\\\\

            \textbf{Example 5}\\
            Task A:\\
            Determine if any two numbers in the given list are closer to each other than a specified threshold.\\
            Task B:\\
            Given an integer array \texttt{nums}, count the elements that have both a strictly smaller and a strictly greater element in the array.\\
            Answer: D\\\\

            \promptheader{Input Tasks}\\
            Task A. \{description1\}\\
            Task B. \{description2\}\\\\

            \promptheader{Output Requirements}\\
            Format your answer exactly as follows:\\
            Answer: [A, B, C, or D]\\
        \end{minipage}\\
        \bottomrule
    \end{tabular}
    \end{adjustbox}
\end{table*}
\begin{table}[t]
    \centering
    \small
    \caption{Prompt for the final screening stage.}
    \label{tab:final-screening-prompt}
    \begin{tabular}{>{\raggedright\arraybackslash\ttfamily}p{0.98\linewidth}}
        \toprule
            \headercolorlong
            \textbf{Instruction}\\
            You will be shown one task description. Your job is to assess whether it describes a basic helper function.\\\\

            \headercolorlong
            \textbf{Definition}\\
            A basic helper function is:\\
            1. Primitive and atomic --- performs a single, irreducible operation.\\
            2. Scalar/boolean output --- returns only a simple scalar or trivial boolean (not a composite structure).\\
            3. Built-in equivalent --- typically maps to a single built-in or standard library function (e.g., abs(x), len(list), max(array)).\\
            4. Subroutine nature --- commonly used as a small sub-step inside larger algorithms.\\\\

            \headercolorlong
            \textbf{Litmus Tests (all must be satisfied for ``Yes'')}\\
            - Built-in mapping: Does it directly correspond to a built-in/standard library call?\\
            - Subroutine usage: Is it normally a utility step within larger problems?\\
            - Atomic simplicity: Does it avoid extra selection, indexing, or multi-step logic?\\\\

            \headercolorlong
            \textbf{Decision Rule}\\
            - Yes: All three tests pass.\\
            - No: Any test fails.\\\\

            \headercolorlong
            \textbf{Output Requirements}\\
            Format your answer exactly as follows:\\

            Decision: Yes | No\\
            Reasoning: (3--4 sentences explaining which tests pass or fail, focusing on atomic simplicity, built-in mapping, and subroutine usage.)\\\\

            \headercolorlong
            \textbf{Task}\\
            \{task description\}\\
        \bottomrule
    \end{tabular}
\end{table}

\section{Instruction Normalization Analysis}
\label{app:normalization-analysis}

Because \tool relies on LLM-based instruction normalization before semantic triage and verification, an important question is whether this step preserves the original meaning of each programming task. To assess this risk, we conduct a small-scale manual analysis of the normalized task descriptions produced by the four LLM backbones used in our evaluation.

We randomly sample task descriptions from all six datasets used in our experiments, including the three evaluation benchmarks and the three post-training corpora. Specifically, we sample 20 task descriptions from each of \textsc{HumanEval}, \textsc{MBPP}, \textsc{LiveCodeBench}, \textsc{CodeAlpaca}, \textsc{Evol-CodeAlpaca}, and \textsc{Magicoder}, resulting in 120 sampled task descriptions in total. For each sampled task, we apply the instruction normalization prompt in Table~\ref{tab:normalization-prompt} using each of the four LLM backbones: \textsc{GPT-5}, \textsc{Gemini-2.5-Pro}, \textsc{gpt-oss-120b}, and \textsc{Qwen3-14B}. This produces 480 original-normalized task pairs for manual analysis.

Two authors independently inspect each original-normalized pair and judge whether the normalized description preserves the original task semantics. We use a binary annotation scheme: \textit{semantics-preserving} if the normalized description retains the same computational objective, input-output behavior, and key constraints as the original task; and \textit{meaning-altering} if it omits, adds, or changes information that would affect the intended solution. Disagreements are resolved through discussion after computing inter-annotator agreement.

\begin{table*}[t]
\centering
\small
\caption{Manual analysis of whether instruction normalization preserves the original task semantics.}
\label{tab:normalization_analysis}
\begin{tabular}{lcccc}
\toprule
\multirow{2}{*}{\textbf{Dataset}}
& \multicolumn{2}{c}{\cellcolor{dustyblue}\textbf{Open-Sourced LLMs}}
& \multicolumn{2}{c}{\cellcolor{sagegreen}\textbf{Closed-Sourced LLMs}} \\
& \cellcolor{dustyblue}\textbf{gpt-oss-120b}
& \cellcolor{dustyblue}\textbf{Qwen3-14B}
& \cellcolor{sagegreen}\textbf{Gemini-2.5-Pro}
& \cellcolor{sagegreen}\textbf{GPT-5} \\
\midrule
\headercolorlong
\multicolumn{5}{c}{\textit{Evaluation Benchmarks}} \\
\midrule
HumanEval      & 20 / 20 (100.0\%) & 20 / 20 (100.0\%) & 20 / 20 (100.0\%) & 20 / 20 (100.0\%) \\
MBPP           & 20 / 20 (100.0\%) & 20 / 20 (100.0\%) & 20 / 20 (100.0\%) & 20 / 20 (100.0\%) \\
LiveCodeBench  & 19 / 20 (95.0\%)  & 18 / 20 (90.0\%)  & 20 / 20 (100.0\%) & 20 / 20 (100.0\%) \\
\midrule
\headercolorlong
\multicolumn{5}{c}{\textit{Post-Training Corpora}} \\
\midrule
CodeAlpaca      & 19 / 20 (95.0\%)  & 19 / 20 (95.0\%)  & 20 / 20 (100.0\%) & 20 / 20 (100.0\%) \\
Evol-CodeAlpaca & 19 / 20 (95.0\%)  & 17 / 20 (85.0\%)  & 20 / 20 (100.0\%) & 20 / 20 (100.0\%) \\
Magicoder       & 20 / 20 (100.0\%) & 20 / 20 (100.0\%) & 20 / 20 (100.0\%) & 20 / 20 (100.0\%) \\
\midrule
\textbf{Overall}
& \textbf{117 / 120 (97.5\%)}
& \textbf{114 / 120 (95.0\%)}
& \textbf{120 / 120 (100.0\%)}
& \textbf{120 / 120 (100.0\%)} \\
\bottomrule
\end{tabular}
\end{table*}

Table~\ref{tab:normalization_analysis} summarizes the results. Overall, normalization preserves task semantics in the vast majority of cases across all four LLM backbones. The main meaning-altering cases arise from omitted edge-case constraints, oversimplified input-output requirements, or accidental removal of auxiliary conditions in longer task descriptions. These cases are relatively rare, suggesting that instruction normalization primarily reduces superficial variation in phrasing and formatting rather than changing the underlying task. The high inter-annotator agreement further indicates that the manual assessment procedure is reliable.

The two annotators achieve substantial agreement, with Cohen's $\kappa = 0.92$. This agreement level suggests that whether normalization preserves task semantics can be judged consistently under our annotation criteria. We therefore conclude that the normalization stage introduces limited semantic distortion while improving the comparability of task descriptions across heterogeneous benchmarks and post-training datasets.

\section{Instruction Normalization Example}
\label{sec:normalization-example}

Table~\ref{tab:normalization-example} shows an example of instruction normalization for two task descriptions drawn from different sources (HumanEval/61 and Magicoder/72803). Although the original descriptions differ substantially in formatting and verbosity, they specify the same programming task. After normalization, both descriptions are rewritten into a concise and consistent format, making their shared computational objective, input specification, and output requirement easier to compare. This pair is later categorized as Functionally Identical.

\begin{table}[t]
\centering
\small
\caption{Example of instruction normalization for two differently structured descriptions of the same programming task drawn from an evaluation benchmark and a post-training corpus.}
\begin{tabular}{p{0.18\linewidth}p{0.76\linewidth}}
\toprule
\multicolumn{2}{l}{\textbf{Original Task Descriptions}} \\
\midrule
Benchmark task &
\begin{minipage}[t]{0.76\linewidth}
\begin{verbatim}
def correct_bracketing(brackets: str):
    """ brackets is a string of "(" and ")".
    return True if every opening bracket has a corresponding
    closing bracket.

    >>> correct_bracketing("(")
    False
    >>> correct_bracketing("()")
    True
    >>> correct_bracketing("(()())")
    True
    >>> correct_bracketing(")(()")
    False
    """
\end{verbatim}
\end{minipage} \\
\addlinespace
Post-training task &
You are tasked with implementing a function that checks whether a given string is a valid parentheses sequence. A valid parentheses sequence is defined as a string consisting only of ``('' and ``)'' characters, where each open parenthesis must have a corresponding closing parenthesis and the pairs must be properly nested.

You need to implement the following function: \texttt{valid\_parentheses(s: str) -> bool}. The function should take a string \texttt{s} as input and return \texttt{True} if the input string is a valid parentheses sequence, and \texttt{False} otherwise. For example, ``(())'' and ``()()'' should return \texttt{True}, while ``)('' should return \texttt{False}. \\
\midrule
\multicolumn{2}{l}{\textbf{Normalized Task Descriptions}} \\
\midrule
Benchmark task &
Determine if a string of parentheses is correctly bracketed, ensuring every opening parenthesis has a matching closing parenthesis. \\
\addlinespace
Post-training task &
Implement a function that verifies whether a given string is a valid parentheses sequence, containing only ``('' and ``)'' characters with correct nesting and matching pairs. \\
\bottomrule
\end{tabular}
\label{tab:normalization-example}
\end{table}

\section{Benchmark Annotation}
\label{app:benchmark_construction}

Annotators are given the task descriptions and, when necessary, the reference solutions to assess semantic equivalence at the level of computational intent and solution strategy. The initial inter-annotator agreement is $\kappa = 0.81$. Disagreements are resolved through discussion to obtain the final labels.

\section{Datasets and Benchmarks}
\label{app:datasets}

This section provides detailed descriptions of the code generation benchmarks and post-training datasets used in our experiments.

\subsection{Benchmarks}

\paragraph{HumanEval.}
HumanEval is a widely used benchmark for evaluating code generation models on Python programming tasks. Each task consists of a natural language description and a set of unit tests used to assess functional correctness.

\paragraph{MBPP.}
MBPP (Mostly Basic Python Problems) is a benchmark containing a diverse collection of beginner-level Python programming tasks with natural language descriptions and reference solutions. Due to its long-term public availability, MBPP is commonly used in contamination analysis studies.

\paragraph{LiveCodeBench-v6.}
LiveCodeBench is a dynamic benchmark designed to mitigate data contamination by continuously collecting new programming problems over time. Tasks are filtered based on release date, making LiveCodeBench suitable for evaluating models under a dynamic and evolving benchmark setting.

\subsection{Post-Training Datasets}

\paragraph{CodeAlpaca-20k.}
CodeAlpaca-20k is an instruction-tuning dataset consisting of approximately 20,000 programming tasks generated by prompting large language models with code-related instructions derived from public sources. The dataset primarily focuses on single-function Python programming problems and is widely used for supervised fine-tuning of code generation models.

\paragraph{Evol-CodeAlpaca-V1.}
Evol-CodeAlpaca-V1 extends CodeAlpaca by iteratively evolving existing tasks through automatic rewriting and complexity augmentation. This evolution process increases task diversity and difficulty, resulting in a larger and more heterogeneous instruction-tuning dataset.

\paragraph{Magicoder-OSS-Instruct-75K.}
Magicoder-OSS-Instruct-75K is a large-scale instruction-tuning dataset constructed from a mixture of open-source code and synthesized programming instructions. It emphasizes algorithmic reasoning and multi-step problem solving, and is commonly used to train high-performance code LLMs.

\section{Models}
\label{app:models}
This section summarizes the embedding model and LLMs used in our experiments.

\subsection{Embedding Models}

We use the text embedding model to perform large-scale semantic triage in Stage~II of \tool. It encodes normalized task descriptions into fixed-dimensional vector representations for similarity search.

\paragraph{jina-embeddings-v3.}
jina-embeddings-v3 is a large-scale text embedding model designed for semantic search, retrieval, and text matching tasks.

\subsection{Large Language Models}

LLMs are used in Stage~III and Stage~IV of \tool to perform fine-grained semantic categorization and trivial task screening.

\paragraph{GPT-5.}
GPT-5 is a proprietary large language model from OpenAI with strong reasoning and code understanding capabilities.

\paragraph{Gemini-2.5-Pro.}
Gemini-2.5-Pro is a high-capacity reasoning-oriented model developed by Google, designed to handle complex semantic analysis tasks.

\paragraph{gpt-oss-120b.}
gpt-oss-120b is an open-weight large language model with publicly available parameters, enabling local deployment and detailed reasoning analysis.

\paragraph{Qwen3 Family.}
The Qwen3 family includes both dense and mixture-of-experts models at multiple scales, providing a range of trade-offs between inference cost and reasoning performance. We use representative variants from this family in ensemble and cascaded classification settings.

\section{Threshold Tuning Details}
\label{sec:threshold}

This section provides supplementary threshold-tuning curves for \tool and the threshold-based baselines. All thresholds are tuned on the same development split described in Section~\ref{sec:golden_dataset}.

\begin{figure*}[t]
\centering

\begin{subfigure}{0.48\textwidth}
    \centering
    \includegraphics[width=\linewidth]{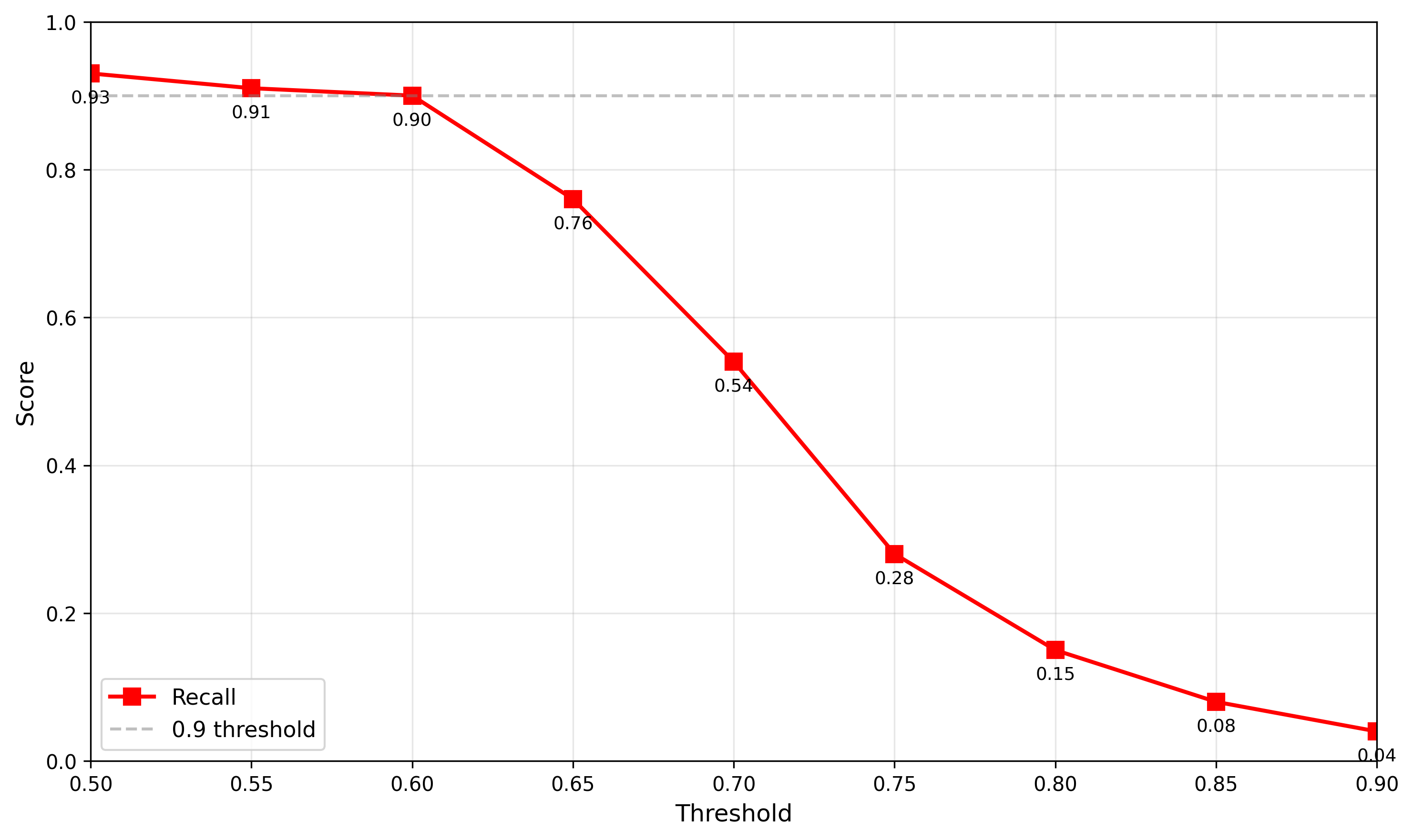}
    \caption{Lower triage threshold $\tau_{\mathrm{low}}$.}
    \label{fig:low}
\end{subfigure}
\hfill
\begin{subfigure}{0.48\textwidth}
    \centering
    \includegraphics[width=\linewidth]{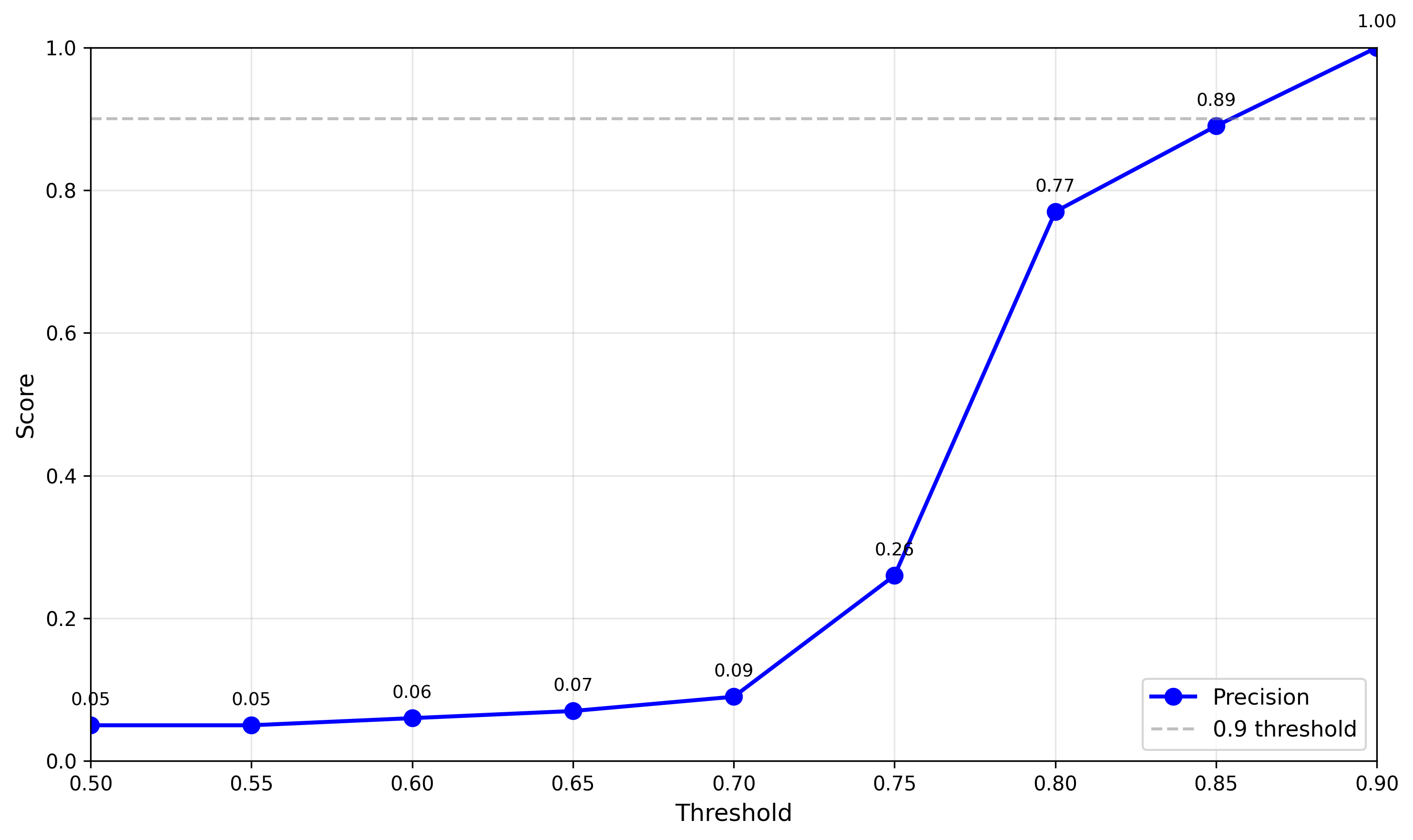}
    \caption{Upper triage threshold $\tau_{\mathrm{high}}$.}
    \label{fig:high}
\end{subfigure}

\vspace{0.75em}

\begin{subfigure}{0.48\textwidth}
    \centering
    \includegraphics[width=\linewidth]{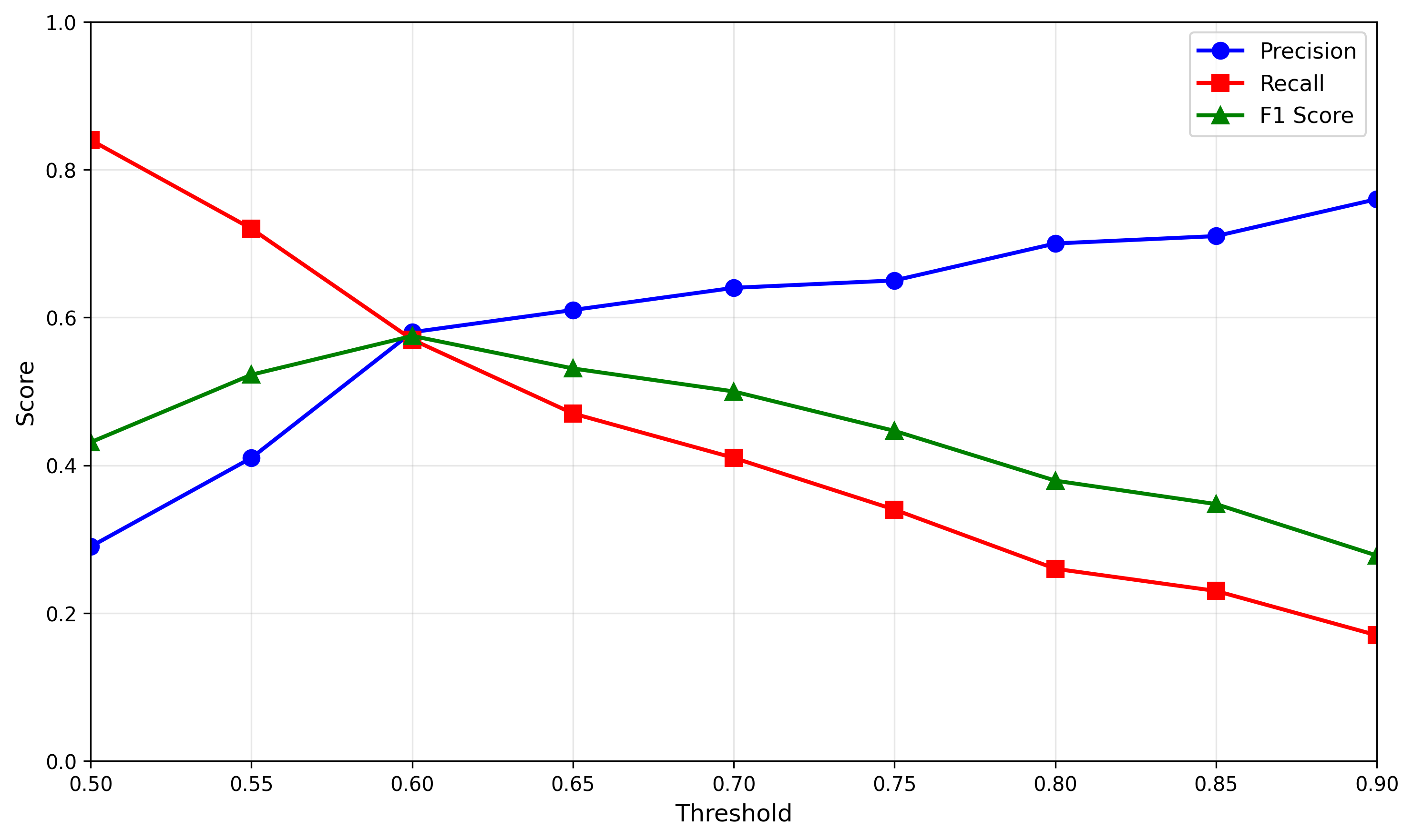}
    \caption{\textsc{Roots} BM25 threshold.}
    \label{fig:roots}
\end{subfigure}
\hfill
\begin{subfigure}{0.48\textwidth}
    \centering
    \includegraphics[width=\linewidth]{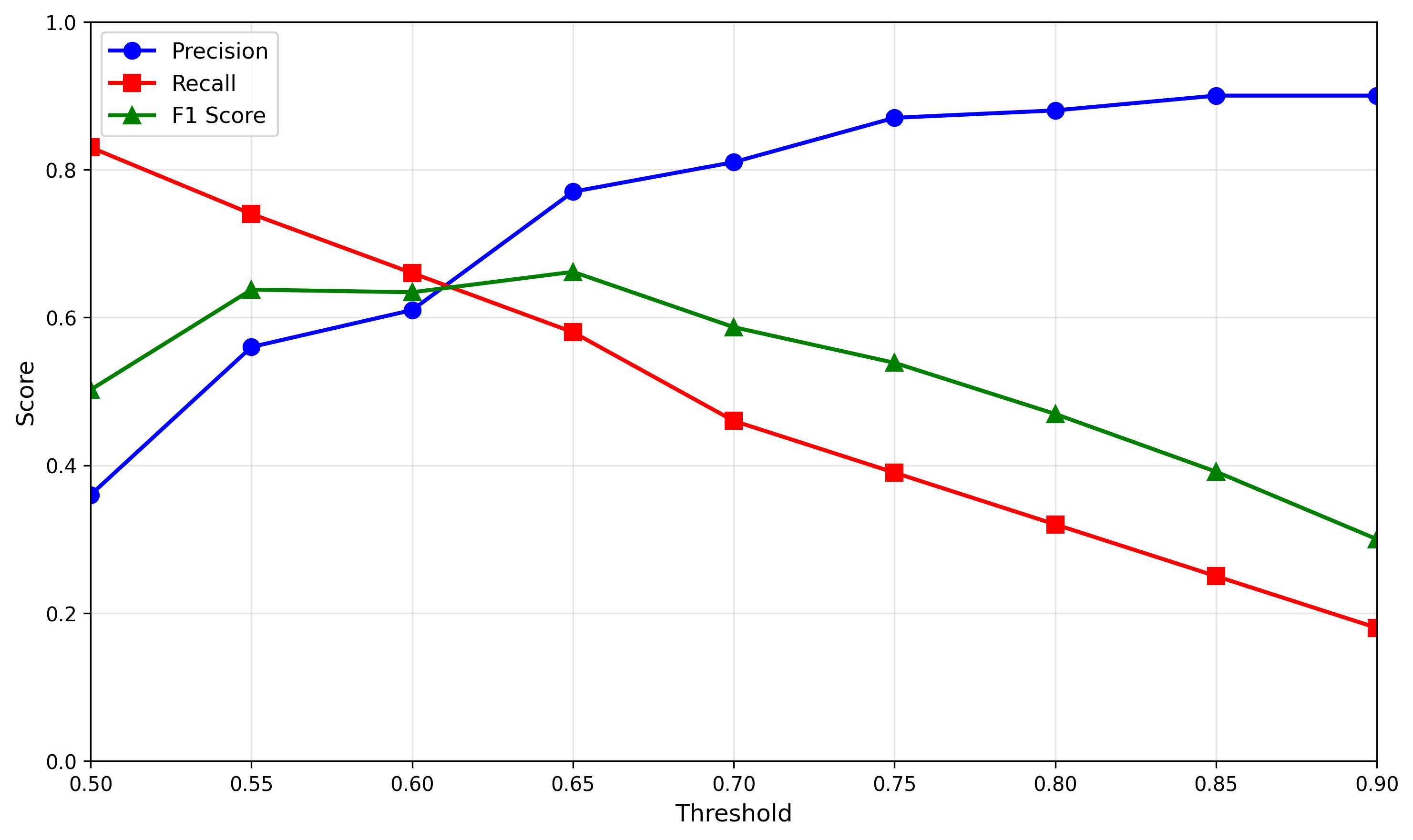}
    \caption{\textsc{PSM} decision threshold.}
    \label{fig:psm}
\end{subfigure}

\vspace{0.75em}

\begin{subfigure}{0.48\textwidth}
    \centering
    \begin{minipage}[c][0.30\textheight][c]{\linewidth}
        \centering
        \includegraphics[width=\linewidth,height=0.29\textheight,keepaspectratio]{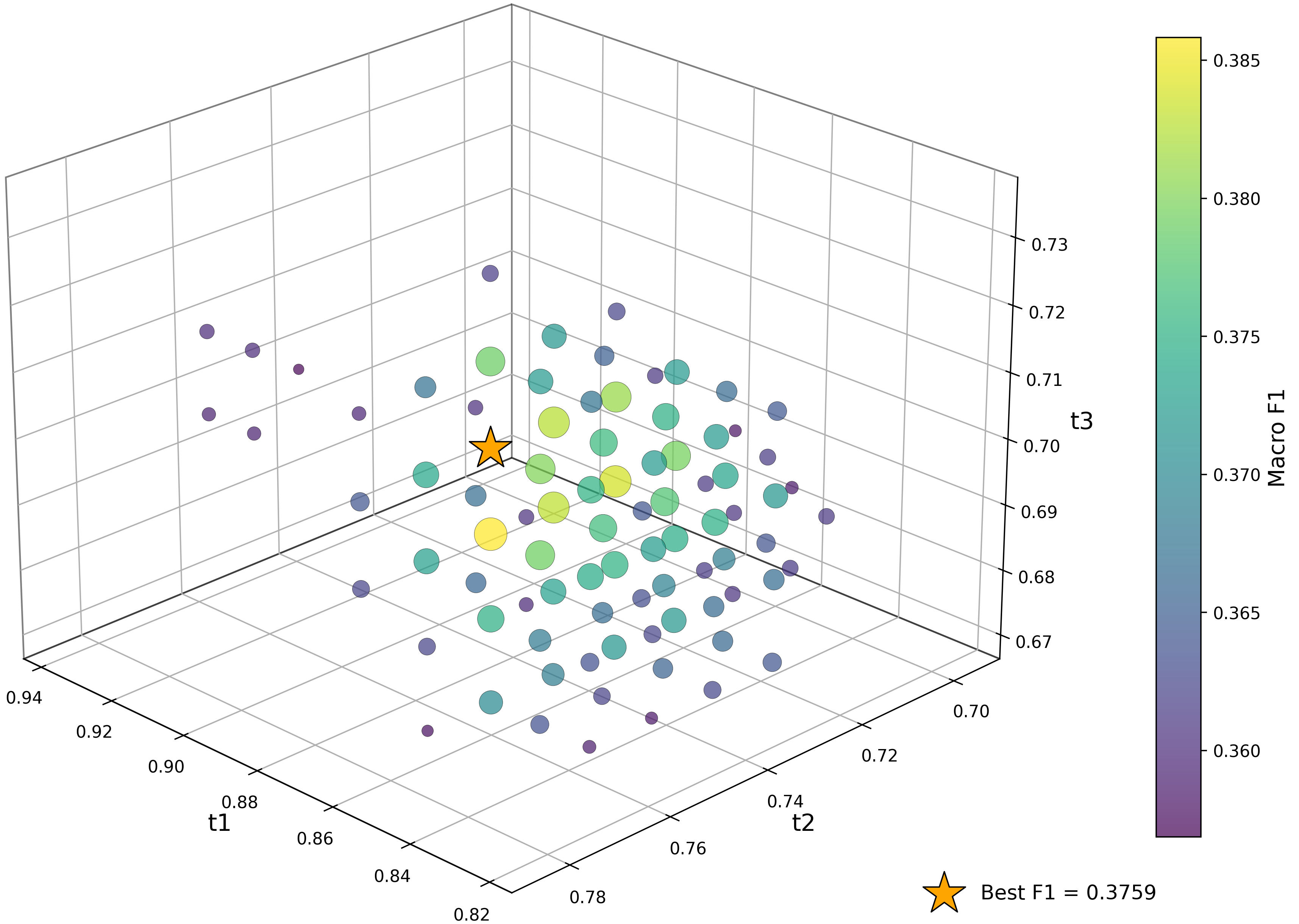}
    \end{minipage}
    \caption{Fine-grained ablation without LLM verification.}
    \label{fig:fine-grained_ablation}
\end{subfigure}
\hfill
\begin{subfigure}{0.48\textwidth}
    \centering
    \begin{minipage}[c][0.30\textheight][c]{\linewidth}
        \centering
        \includegraphics[width=\linewidth,height=0.29\textheight,keepaspectratio]{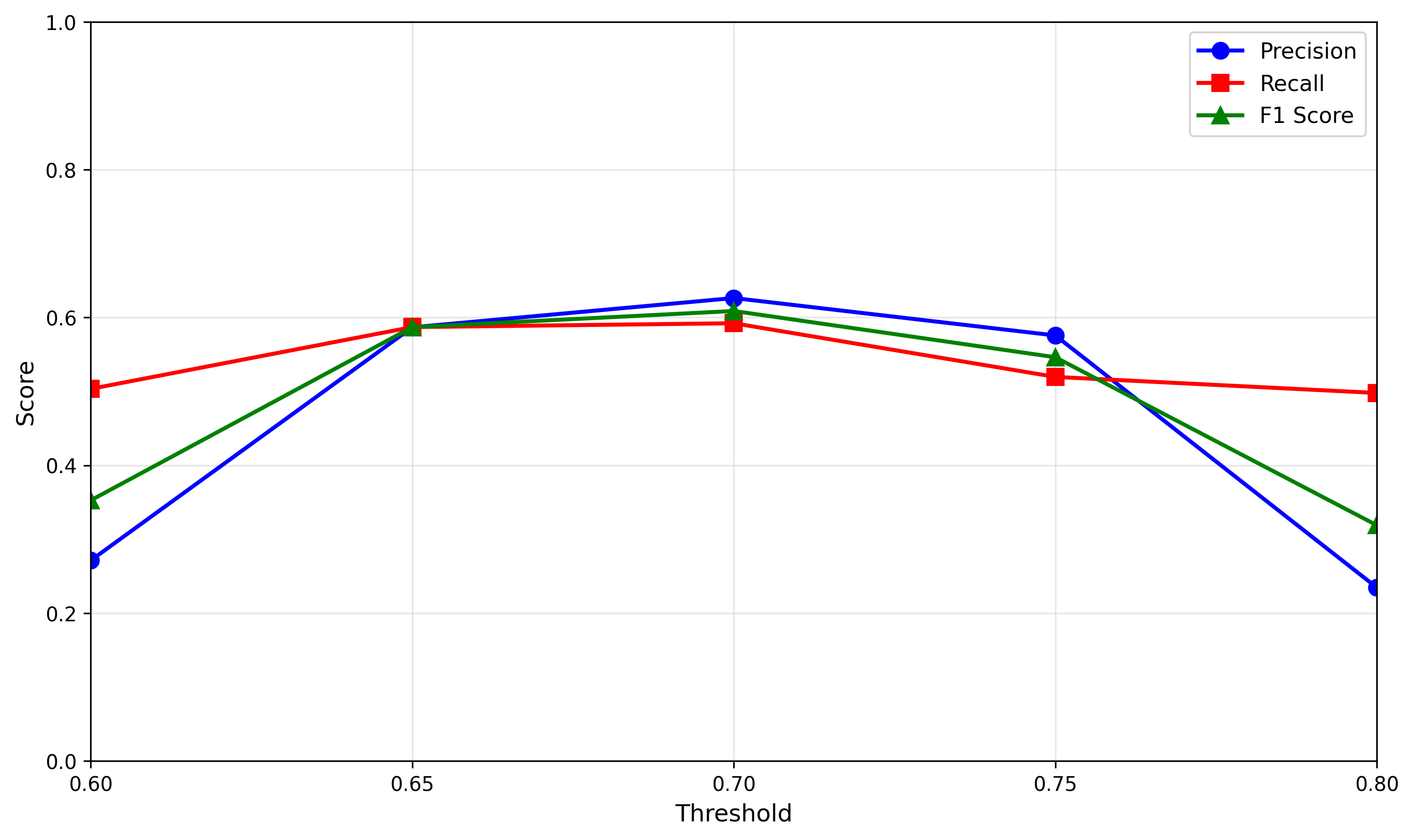}
    \end{minipage}
    \caption{Binary ablation without LLM verification.}
    \label{fig:binary_ablation}
\end{subfigure}

\caption{Threshold sensitivity analyses for triage, baseline methods, and the ablation variants.}
\label{fig:threshold_analysis}
\end{figure*}

For \tool, Figures~\ref{fig:low} and~\ref{fig:high} show the development-set behavior used to select the lower triage threshold $\tau_{\mathrm{low}}$ and the upper triage threshold $\tau_{\mathrm{high}}$, respectively. As described in Section~\ref{sec:experimental-design}, $\tau_{\mathrm{low}}$ is selected to preserve high recall over contaminated pairs before LLM-based verification, while $\tau_{\mathrm{high}}$ is selected to identify highly confident Functionally Identical matches.

For \textsc{Roots} and \textsc{Program Similarity Matching (PSM)}, we tune the BM25 and program-similarity thresholds on the same development split by maximizing binary F1. Figures~\ref{fig:roots} and~\ref{fig:psm} show the corresponding threshold sensitivity curves.

For the ablation that removes LLM-based fine-grained verification, we evaluate two threshold-only variants. The first variant directly performs fine-grained classification using three decision thresholds. Pairs above $t_1$ are classified as Functionally Identical, pairs between $t_1$ and $t_2$ as Nearly Identical, pairs between $t_2$ and $t_3$ as Shared Logic, and pairs below $t_3$ as Unrelated. We jointly tune $t_1$, $t_2$, and $t_3$ on the development split by maximizing macro-F1. Figure~\ref{fig:fine-grained_ablation} visualizes the resulting macro-F1 landscape over the three-threshold search space, with the best configuration achieved at $t_1 = 0.86$, $t_2 = 0.75$, and $t_3 = 0.71$. The second variant follows the binary reduction used in Section~\ref{sec:ablation}, where Functionally Identical, Nearly Identical, and Shared Logic are grouped as contaminated and Unrelated is treated as non-contaminated. For this binary variant, we tune a single decision threshold on the development split by maximizing binary F1. Figure~\ref{fig:binary_ablation} reports the corresponding threshold sensitivity analyses.

\section{Failure Cases}
\label{sec:failure-cases}

To further analyze the limitations of \tool, we examine the 59 misclassified pairs produced by \textsc{GPT-5} on the evaluation benchmark. We identify four recurring error types.

\subsection{Core Logic Misidentification (50.85\%)}

\begin{table}[t]
    \centering
    \small
    \caption{Example of misclassified pairs caused by core logic misidentification.}
    \label{tab:error-type-1}
    \renewcommand{\arraystretch}{1.15}
    \begin{tabular}{p{0.96\linewidth}}
        \toprule
        \textbf{Task Description 1} \\
        \midrule
        Return a list of prime factors of the given integer, ordered from smallest to largest, with each factor appearing as many times as it does in the factorization. The product of the factors should equal the input number. \\
        
        \midrule
        \textbf{Task Description 2} \\
        \midrule
        Implement a PHP function that accepts an integer and returns its prime factorization as a list of integers. \\
        
        \midrule
        \textbf{Reasoning Summary} \\
        \midrule

        I need to analyze the relationship between two tasks: Task A is about returning a list of prime factors of an integer, ordered from smallest to largest, with each factor appearing as many times as it does in the factorization. Task B is about implementing a PHP function that accepts an integer and returns its prime factorization as a list of integers. Both tasks are fundamentally about prime factorization but differ in implementation language and output format. The core algorithm is the same, so I would classify this as a similar task as option B. \\
        
        \midrule
        \textbf{Detection Result} \\
        \midrule
        Nearly Identical (B) \\
        
        \midrule
        \textbf{Ground Truth} \\
        \midrule
        Functionally Identical (A) \\
        \bottomrule
    \end{tabular}
\end{table}

Core logic misidentification is the most common failure mode, accounting for 30 out of 59 misclassifications (50.85\%). In these cases, the LLM backbone fails to abstract the shared algorithmic idea between two tasks and instead relies too heavily on surface framing, local constraints, or implementation details. As a result, semantically related task pairs may be judged as less related than they actually are. This suggests that a central challenge in fine-grained contamination detection is not merely recognizing topical similarity, but correctly identifying the deeper computational structure shared by two task descriptions. Table~\ref{tab:error-type-1} presents a representative example.

\subsection{Confusion Between Adjacent Semantic Categories (37.29\%)}

\begin{table}[t]
    \centering
    \small
    \caption{Example of a misclassified pair caused by confusion between adjacent semantic categories.}
    \label{tab:error-type-2}
    \renewcommand{\arraystretch}{1.15}
    \begin{tabular}{p{0.96\linewidth}}
        \toprule
        \textbf{Task Description 1} \\
        \midrule
        Filter a given list to include only integers. \\
        
        \midrule
        \textbf{Task Description 2} \\
        \midrule
        Implement a program that filters out all odd numbers from a given array of integers. \\
        
        \midrule
        \textbf{Reasoning Summary} \\
        \midrule

        I’m considering two tasks: Task A is about filtering a list to only include integers, while Task B filters out odd numbers and keeps only even integers. They share a filtering goal but use different predicates -- one is a type check and the other focuses on parity. It seems that while both are filtering, they are not using the same algorithmic method. Task C suggests they use a similar filtering logic with differing problems, so I’d lean towards C. \\
        
        \midrule
        \textbf{Detection Result} \\
        \midrule
        Shared Logic (C) \\
        
        \midrule
        \textbf{Ground Truth} \\
        \midrule
        Unrelated (D) \\
        \bottomrule
    \end{tabular}
\end{table}

The second major error pattern is confusion between adjacent semantic categories, especially between Nearly Identical and Shared Logic. This category accounts for 22 out of 59 misclassifications (37.29\%). These cases typically arise when two tasks share substantial problem structure but differ in a limited number of constraints, assumptions, or output requirements. In such settings, the LLM backbone often recognizes that the tasks are related, but assigns the wrong fine-grained label because the boundary between neighboring categories is inherently subtle. Table~\ref{tab:error-type-2} shows a representative example.

\subsection{Hallucination or Internally Inconsistent Reasoning (8.47\%)}

\begin{table}[t]
    \centering
    \small
    \caption{Example of a misclassified pair caused by hallucination or internally inconsistent reasoning.}
    \label{tab:error-type-3}
    \renewcommand{\arraystretch}{1.15}
    \begin{tabular}{p{0.96\linewidth}}
        \toprule
        \textbf{Task Description 1} \\
        \midrule
        Determine if a given string is a palindrome. Then, generate the shortest palindrome starting with the provided string by appending the reverse of its prefix that precedes the longest palindromic suffix. \\
        
        \midrule
        \textbf{Task Description 2} \\
        \midrule
        Implement a Python program that checks if a given phrase is a palindrome by removing spaces, ignoring case, and verifying if the phrase reads the same forward and backward. \\
        
        \midrule
        \textbf{Reasoning Summary} \\
        \midrule

        I need to tackle two tasks related to palindromes. The first is to check if a string is a palindrome and then generate the shortest palindrome by appending the reverse of its prefix that comes before the longest palindromic suffix. Essentially, I’ll start with the given string and ensure it forms a palindrome. The second task is to create a Python program that checks if a given phrase is a palindrome by ignoring spaces and case sensitivity. \\
        
        \midrule
        \textbf{Detection Result} \\
        \midrule
        Unrelated (D) \\
        
        \midrule
        \textbf{Ground Truth} \\
        \midrule
        Shared Logic (C) \\
        \bottomrule
    \end{tabular}
\end{table}

A smaller portion of errors, 5 out of 59 cases (8.47\%), are caused by hallucinated or internally inconsistent reasoning. In these cases, the LLM backbone introduces functionality or task properties that are not supported by the input descriptions, or produces reasoning that does not align with the final predicted label. Although less frequent, this failure mode reflects a limitation of LLM-based verification itself: even when relevant evidence is present, the reasoning process may still become unstable or partially unsupported. Table~\ref{tab:error-type-3} provides an example.

\subsection{Formatting Artifact Bias (3.39\%)}

\begin{table}[t]
    \centering
    \small
    \caption{Example of a misclassified pair caused by formatting artifact bias.}
    \label{tab:error-type-4}
    \renewcommand{\arraystretch}{1.15}
    \begin{tabular}{p{0.96\linewidth}}
        \toprule
        \textbf{Task Description 1} \\
        \midrule
        Given a string representing a file name, write a function to determine if the name is valid based on specific criteria. The function should return 'Yes' if the name meets all the conditions, and 'No' otherwise. Conditions include no more than three digits, exactly one dot, a non-empty substring before the dot starting with a letter, and a valid extension ('txt', 'exe', or 'dll'). \\
        
        \midrule
        \textbf{Task Description 2} \\
        \midrule
        Implement a function to validate an email address using a regular expression. \\
        
        \midrule
        \textbf{Detection Result} \\
        \midrule
        Unrelated (D) \\
        
        \midrule
        \textbf{Ground Truth} \\
        \midrule
        Shared Logic (C) \\
        \bottomrule
    \end{tabular}
\end{table}

The remaining 2 out of 59 errors (3.39\%) stem from persistent formatting artifacts that survive our normalization procedure. These surface-level similarities can mislead the LLM backbone into assigning overly strong relatedness to task pairs whose underlying computational objectives are in fact different. This pattern highlights a limitation of the current normalization and filtering process: incidental stylistic resemblance can still be mistaken for meaningful algorithmic overlap. Table~\ref{tab:error-type-4} presents a representative example.

\clearpage



\end{document}